\documentclass[pre,twocolumn]{revtex4-2}
\usepackage{amsmath}
\usepackage{amssymb}
\usepackage{graphicx}
\usepackage{braket}
\usepackage{stmaryrd}
\usepackage{hyperref}
\usepackage{color}
\usepackage{dsfont}
\usepackage{bm}
\usepackage{wasysym}
\usepackage{booktabs}

\hypersetup{
    colorlinks,
    citecolor=blue,
    linkcolor=blue,
    urlcolor=blue
}

\newcounter{EMappendix}
\renewcommand{\theEMappendix}{\Alph{EMappendix}}
\newcommand{\EMsection}[1]
{
\stepcounter{EMappendix}
\textit{Appendix \theEMappendix: #1.---}
}

\newcommand{\dummyblock}[1]{}

\begin{document}

\title{
Thermodynamics of Classifiers
}
\author{Yoshihiko Hasegawa}
\email{hasegawa@biom.t.u-tokyo.ac.jp}
\affiliation{Department of Electrical Engineering and Information Systems, Graduate School of Engineering, The University of Tokyo,
Tokyo 113-8656, Japan}

\date{\today}
\begin{abstract}

The rise of artificial intelligence has led to a growing demand for computational power.
In this situation, energy-efficient computing methods are being actively investigated, which raises a fundamental question: what is the relationship between computational error and thermodynamic cost?
Here, we derive the error-cost trade-off in the binary classifier by considering classification based on nonequilibrium thermodynamics.
We derive trade-off relations showing that classification error cannot be lower than fundamental bounds set by thermodynamic quantities: entropy production, which measures irreversibility, and dynamical activity, which quantifies the activity of the system. 
Our results show that when entropy production or dynamical activity vanishes, the classification error reaches $1/2$, equivalent to random guessing, whereas greater thermodynamic costs enable lower error. 
The results establish a quantitative trade-off in physical information processing, potentially leading to the development of energy-efficient computing devices.

\end{abstract}
\maketitle

\textit{Introduction.---}Advances in artificial intelligence are bringing about significant transformations in society. However, behind these developments, the growing computational energy consumption has become a serious problem.
This has created a strong demand for developing ``thermodynamics of intelligence'', which is expected to establish a fundamental relationship between computational error and thermodynamic cost.
Revealing the connection between information and thermodynamics has a long history. 
Since the discovery of the Landauer principle \cite{Landauer:1961:LP,Bennett:2003:Landauer}, it has been recognized that irreversible information processing inevitably requires thermodynamic costs.
Since then, numerous studies have investigated the thermodynamic costs of various information-processing systems.
Bennett~\cite{Bennett:1973:LogicalReversibility} showed that reversible computation can achieve arbitrarily low dissipation, and 
Zurek~\cite{Zurek:1989:ThermodynamicCostComputation} extended the Landauer principle to incorporate Kolmogorov complexity.
Recent advances in stochastic thermodynamics \cite{Seifert:2012:FTReview} have enabled a rigorous formulation of the thermodynamic costs of computation \cite{Sagawa:2013:InformationProcessing}.
References~\cite{Kolchinsky:2017:DissipationInitialDistribution,Wolpert:2020:ThermodynamicsComputingCircuits,Manzano:2024:AbsoluteIrreversibility} considered entropy production as a consequence of using different initial distributions, termed the mismatch cost.
The thermodynamic cost of learning \cite{Goldt:2017:STLearning} was formulated based on the thermodynamics of information flow \cite{Horowitz:2014:InfoFlow}. 
Reference~\cite{Kolchinsky:2020:ThermodynamicCosts} formulated the thermodynamic cost in Turing machines and indicated the relation between complexity and entropy production. 
More recently, Ref.~\cite{Ikeda:2025:SpeedAccuracyDiffusion} examined the speed-accuracy trade-off in diffusion models.
For a comprehensive review in this direction, please see Refs.~\cite{Wolpert:2019:StochasticComputation,Wolpert:2024:stochastic-thermodynamics-computation}.

Here, we reveal the relationship between the thermodynamic cost and the error of classifiers. Given data, classifiers are machines that predict the underlying class label. 
We consider classifiers constructed from Markov processes and show that the misclassification error of such classifiers is bounded from below by thermodynamic costs, such as entropy production and dynamical activity [Eqs.~\eqref{eq:result1}, \eqref{eq:result2}, and \eqref{eq:result_da1}].
Our central finding is an error-cost trade-off: when entropy production or dynamical activity vanishes, the error equals $1/2$, no better than random guessing, and it can fall below this value only as the thermodynamic costs grow. 
From the information-processing perspective,
Markov-process classification is widely used in time-series prediction and recognition \cite{Rabiner:1989:hmm-tutorial,Smyth:1997:hmm-clustering,Bishop:2006:PRML,Esling:2012:TimeSeriesDataMining,Panuccio:2002:sequential-clustering,Stella:2012:ctbn-classifiers,Spaeh:2024:markov-chain-mixtures}.
From the physical perspective,
it can be used to model cellular information processing, particularly as a framework for understanding how biochemical receptors discriminate between ligands.
Pioneering works \cite{Hopfield:1974:KineticProofreading,Ninio:1975:KineticAmplification} proposed a kinetic proofreading model and showed how energy-driven nonequilibrium processes can significantly improve ligand specificity, and later research expanded this concept based on nonequilibrium thermodynamics 
\cite{McKeithan:1995:KineticProofreading,Murugan:2014:DiscriminatoryProofreading,Hartich:2015:NonequilibriumSensing,Rao:2015:ThermodynamicsAccuracyProofreading,Wong:2018:EnergySpeedAccuracy,Ganti:2020:TCellSignaling,Berx:2024:tradeoffs-thermodynamics-energy-relay}. Another related topic is concentration sensing, where cellular receptors estimate the ligand concentration
\cite{Berg:1977:PhysicsOfChemoreception,Endres:2009:MLE,Mora:2010:MLE,Mehta:2012:Energetic,Govern:2014:EnergyDissipationNoise,Harvey:2023:energy-accuracy-tradeoffs}. 
Because these previous studies are based on statistical methods, divergences and the Fisher information play important roles. However, because it is generally difficult to express divergence and Fisher information (or their lower and upper bounds) in terms of physical quantities, specific physical models are often assumed for the underlying mechanisms.
In this study, we express an upper bound of the divergence between general distributions in terms of entropy production and dynamical activity.
The derivation is based on considering independent Markov processes and a coupled measurement, which is reminiscent of multi-copy measurement in quantum information \cite{Peres:1991:OptimalDetection,Massar:1995:OptimalExtraction,Gill:2000:StateEstimation,Ekert:2002:DirectEstimations,Acin:2005:MultipleCopyDiscrimination}. 
Because of this generality, 
our model does not assume how transition rates are affected across classes, and thus
the derived bounds hold in general classification settings.

\begin{figure}
\centering
\includegraphics[width=1\linewidth]{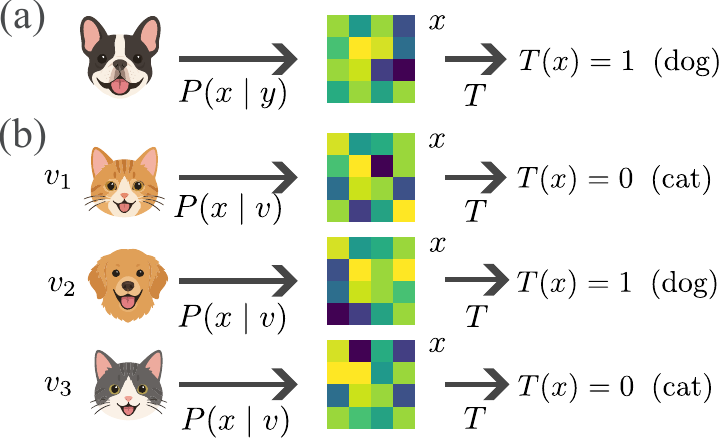}
\caption{
Classification scenarios. (a) Classification of binary data $y=0$ (cat) and $y=1$ (dog). 
Given $y \in \{0,1\}$, a stochastic machine outputs $x$, which is a multidimensional vector. The mapping function $T(x)$ yields $0$ or $1$, which is a prediction made by the classifier. 
The difference between $y$ and $T(x)$ is the error. 
(b) Classification of multiple data points $\mathcal{V} = \{v_1,v_2,v_3\}$. 
A data point $v$ is either a dog or a cat, and we do not specify types of cat or dog.
Each data point has a true class label denoted by $L(v) \in \{0,1\}$ (i.e., $L(v)$ is dog or cat).
Given a data point $v$, a stochastic machine outputs $x$. The mapping function $T(x)$ yields $0$ or $1$, which is a prediction made by the classifier. 
The difference between $L(v)$ and $T(x)$ is the error for the data point $v$.
}
\label{fig:classification_scenario}
\end{figure}

\textit{Methods.---}
We consider a classification task based on stochastic thermodynamics. We consider a continuous-time Markov process on a state space $\mathfrak{B}=\{1,2,\cdots,D\}$.
Let the state of the system at time \(t \ge 0\) be represented by $z(t) \in \mathfrak{B}$.
We define the transition rate from state ${\mu}$ to state ${\nu}$ as $W_{\nu\mu}$.
We denote the probability of being in state ${\mu}$ at time $t$ by
$P(\mu;t)$. 
Then the time evolution of the probability distribution follows the master equation:
\begin{align}
    \frac{d}{dt}P(\nu;t)=\sum_{\mu}W_{\nu\mu}P(\mu;t),
    \label{eq:master_equation}
\end{align}
where the diagonal elements are defined by $W_{\mu\mu}\equiv-\sum_{\nu\neq\mu}W_{\nu\mu}$.
Here, we assume that the transition rate $W_{\nu\mu}$ is time-independent. 
We define relevant thermodynamic quantities associated with the Markov process.
Assuming the local detailed balance condition, 
the time-integrated entropy production is given by
\begin{align}
    \Sigma(\tau)&\equiv\int^{\tau}_{0}\sum_{\mu\ne\nu}W_{\mu\nu}P(\nu;t)\ln\frac{W_{\mu\nu}P(\nu;t)}{W_{\nu\mu}P(\mu;t)}dt.
    \label{eq:EP_def}
\end{align}
The entropy production comprises contributions from heat dissipation and the change in the Shannon entropy of the state.
The entropy production is a central quantity in nonequilibrium thermodynamics: the trajectory entropy production satisfies the integrated fluctuation theorem \cite{Seifert:2005:FT} and serves as a lower bound in thermodynamic uncertainty relations \cite{Barato:2015:UncRel,Gingrich:2016:TUP}.
The time-integrated dynamical activity is defined by
\begin{align}
    \mathcal{A}(\tau)\equiv\int^{\tau}_{0}\mathfrak{a}(t)dt,
    \label{eq:integrated_activity_def}
\end{align}
where $\mathfrak{a}(t)$ is the dynamical activity:
\begin{align}
    \mathfrak{a}(t)\equiv\sum_{\mu\neq\nu}W_{\mu\nu}P(\nu;t).
    \label{eq:activity_def}
\end{align}
Dynamical activity measures the average number of jump events occurring within the time interval $[0,\tau]$. This metric captures the intensity of system dynamics and plays a crucial role in stochastic thermodynamics \cite{Garrahan:2017:TUR,Shiraishi:2018:SpeedLimit,Terlizzi:2019:KUR,Hasegawa:2020:QTURPRL,Hasegawa:2023:BulkBoundaryBoundNC}.

\textit{Results.---}
We introduce the classification scenario considered in this study. 
The variable $y$ represents the object to be classified.
In the case considered here, $y$ has two possible values: $0$ or $1$.
We consider a stochastic machine that takes $y$ as input and outputs $x$, where the conditional probability is expressed by $P(x \mid y)$. 
Here, the output $x$ is not necessarily binary.
It can be a multi-dimensional vector or even infinite-dimensional data. 
We apply the mapping function $T(x)$ (such as the logistic function and thresholding) to reduce the output $x$ to binary data, $T(x)\in \{0,1\}$.
We regard $T(x)$ as a prediction of the class label of $y$ by the stochastic classifier. See Fig.~\ref{fig:classification_scenario}(a) for an illustrative example. 
We are interested in the misclassification error.
Given $y$, the misclassification probability reads
\begin{align}
    P_{\mathrm{err}}(y)\equiv \sum_{x}\mathds{1}\{T(x)\ne y\}P(x\mid y),
    \label{eq:Perr_y_def}
\end{align}
where $\mathds{1}\{\cdots\}$ is the indicator function. 
Suppose that the two types of data ($y=0$ or $1$) are prepared according to the prior probability $\pi_y$. 
The average error is
\begin{align}
    P_{\mathrm{err}}^{\mathrm{ave}}\equiv \sum_{y}P_{\mathrm{err}}(y)\pi_{y}.
    \label{eq:Perr_Bayes_error}
\end{align}
It is known that $P_\mathrm{err}^\mathrm{ave}$ attains its minimum when $T(x)$ is given by
\begin{align}
    T(x)=\underset{y}{\mathrm{argmax}}\,P(y\mid x).
    \label{eq:Bayes_classifier_def}
\end{align}
The average error for $T(x)$ in Eq.~\eqref{eq:Bayes_classifier_def} is given by \cite{Devroye:1996:pattern-recognition,Nielsen:2014:Generalized}
\begin{align}
P_\mathrm{err}^\mathrm{Bayes} &\equiv \min_{T(x)}P^{\mathrm{ave}}_{\mathrm{err}}
=\sum_{x}\min_{y}\left[P(x\mid y)\pi_{y}\right]\nonumber\\&=\frac{1}{2}-\mathrm{TV}\left(P(x\mid y=0)\pi_{0},P(x\mid y=1)\pi_{1}\right),
    \label{eq:Bayes_error_classical}
\end{align}
where $\mathrm{TV}(\cdot,\cdot)$ is the total variation distance.
Equation~\eqref{eq:Bayes_error_classical} is known as the Bayes error. 
The average error cannot fall below Eq.~\eqref{eq:Bayes_error_classical} for any $T(x)$, i.e., $P_\mathrm{err}^\mathrm{ave}\ge P_\mathrm{err}^\mathrm{Bayes}$.

We now present an explicit classification process based on the Markov process for $P(x\mid y)$ in Eq.~\eqref{eq:Perr_y_def}. 
We assume that the transition rate $W_{\nu\mu}$ depends on the underlying class $y$, while the state space is shared by both classes $y=0$ and $y=1$.
For the label $y\in\{0,1\}$, the transition rate is expressed as $W^{(y)}_{\nu\mu}$.
Therefore, the dimensions of $W^{(0)}_{\nu\mu}$ and $W^{(1)}_{\nu\mu}$ are the same. 
The probability $P(\mu;t)$ and state $z(t)$ corresponding to the Markov process of class $y$ are denoted by $P^{(y)}(\mu;t)$ and $z^{(y)}(t)$, respectively.
We assume that the initial state of the process is identical for the two classes, $z^{(0)}(0) = z^{(1)}(0) \in \mathfrak{B}$.
Given the final state $z^{(y)}(\tau)$, the mapping $T(x)=T(z(\tau))$ yields the prediction of the underlying class label. 
Therefore, $x$ in Eq.~\eqref{eq:Perr_y_def} corresponds to $x=z(\tau)$ in this setting. 
For example, in cellular ligand discrimination, $y$ corresponds to the ligand type (e.g., 0: harmless ligand, 1: harmful ligand), and a cellular receptor classifies the ligand based on stochastic biochemical reactions.
It is often modeled that, when different ligands bind to the receptor, the transition rates of the underlying biochemical reactions are different for each ligand \cite{Hopfield:1974:KineticProofreading,Ninio:1975:KineticAmplification,McKeithan:1995:KineticProofreading,Murugan:2014:DiscriminatoryProofreading, Wong:2018:EnergySpeedAccuracy,Ganti:2020:TCellSignaling}. 
Considering different transition rates $W_{\nu\mu}^{(y)}$ for each class, as modeled in our study, properly reflects this ligand discrimination model.

In binary classification based on the final state $z(\tau)$ of the Markov process, we derive lower bounds on the average error $P_\mathrm{err}^\mathrm{ave}$.
For simplicity, we assume uniform priors, $\pi_0=\pi_1=1/2$, hereafter.
However, it is straightforward to generalize to arbitrary priors (see Supplementary Material \cite{Supp:2026:ThermoClassifier}).
Specifically, the following relation holds:
\begin{align}
    P^{\mathrm{ave}}_{\mathrm{err}}\geq\frac{1}{2}\left[1-\sin\left(\frac{\int^{\tau}_{0}\frac{\sqrt{\Sigma^{(0)}(t)+\Sigma^{(1)}(t)}}{t}dt}{2\sqrt{2}}\right)\right].
    \label{eq:result1}
\end{align}
Here, $\Sigma^{(0)}(t)$ and $\Sigma^{(1)}(t)$ are the entropy productions generated by the Markov processes of $y=0$ and $y=1$, respectively.
Equation~\eqref{eq:result1} holds within the following range: $0\le(1/(2\sqrt{2}))\int^{\tau}_{0}\sqrt{\Sigma^{(0)}(t)+\Sigma^{(1)}(t)}/t\,dt\le\pi/2$.
Outside this range, Eq.~\eqref{eq:result1} becomes vacuous. 
The proof of Eq.~\eqref{eq:result1} is provided in the End Matter.
The derivation uses a combined Markov process comprising independent processes, which is inspired by multi-copy measurement in quantum information \cite{Peres:1991:OptimalDetection,Massar:1995:OptimalExtraction,Gill:2000:StateEstimation,Ekert:2002:DirectEstimations,Acin:2005:MultipleCopyDiscrimination}.
Equation~\eqref{eq:result1} represents a lower bound on the average error, which provides an ultimate error bound for classification.
Therefore, the sum of the costs to drive each of the two classes defines the lower bound on the average error.
When this entropy production vanishes, $\Sigma^{(0)}(\tau)+\Sigma^{(1)}(\tau)=0$, the right-hand side of Eq.~\eqref{eq:result1} becomes $1/2$, which is the error of random guessing and indicates that the classifier cannot predict the class label reliably.
Higher entropy production $\Sigma^{(0)}(\tau)+\Sigma^{(1)}(\tau)$ yields a smaller lower bound, which provides an energy-error trade-off in binary classification.

Equation~\eqref{eq:result1} provides a lower bound on the average error in terms of the entropy production. 
Another thermodynamic cost, the time-integrated dynamical activity $\mathcal{A}(t)$ [Eq.~\eqref{eq:integrated_activity_def}], also constitutes a lower bound on the average error.
Specifically, via a calculation similar to that for Eq.~\eqref{eq:result1}, the following inequality holds:
\begin{align}
    P^{\mathrm{ave}}_{\mathrm{err}}\ge\frac{1}{2}\left[1-\sin\left(\frac{\int^{\tau}_{0}\frac{\sqrt{\mathcal{A}^{(0)}(t)+\mathcal{A}^{(1)}(t)}}{t}dt}{2}\right)\right].
    \label{eq:result2}
\end{align}
Here, $\mathcal{A}^{(0)}(t)$ and $\mathcal{A}^{(1)}(t)$ are the dynamical activities of the Markov processes of $y=0$ and $y=1$, respectively.
Again, the following condition should be met in Eq.~\eqref{eq:result2}: $0\le(1/2)\int^{\tau}_{0}\sqrt{\mathcal{A}^{(0)}(t)+\mathcal{A}^{(1)}(t)}/t\,dt\le\pi/2$. 
The derivation is shown in the End Matter. 
Equation~\eqref{eq:result2} represents the average error bound based on dynamical activity.
Equation~\eqref{eq:result2} shows that the sum of the dynamical activities for each of the two classes defines the lower bound on the average error.
This result is intuitive. Without dynamics, $\mathcal{A}(\tau) = 0$, nothing changes, making classification impossible. Greater dynamical activity, on the other hand, allows the two cases to exhibit more distinct behavior, making it easier for a classifier to predict the underlying class label.

Using a different approach, we can obtain an alternative bound for the average error in terms of the dynamical activity: 
\begin{align}
    P^{\mathrm{ave}}_{\mathrm{err}}\ge\frac{1}{2}\left(1-\sqrt{1-e^{-\tau\left[\mathfrak{a}^{(0)}(0)+\mathfrak{a}^{(1)}(0)\right]}}\right),
    \label{eq:result_da1}
\end{align}
Here, $\mathfrak{a}^{(0)}(t)$ and $\mathfrak{a}^{(1)}(t)$ are the dynamical activities [Eq.~\eqref{eq:activity_def}] of the Markov processes corresponding to $y=0$ and $y=1$, respectively. 
The derivation is shown in the End Matter.
Equation~\eqref{eq:result_da1} shows that the sum of the dynamical activities at time $t=0$ is key for classification.
Unlike Eq.~\eqref{eq:result2}, which becomes vacuous for larger dynamical activity, Eq.~\eqref{eq:result_da1} holds for any value of $\mathfrak{a}^{(0)}(0)+\mathfrak{a}^{(1)}(0)$. 
However, for transient dynamics, it is unclear whether Eq.~\eqref{eq:result2} or Eq.~\eqref{eq:result_da1} is tighter (note that for state-based classification, we cannot expect a steady-state situation).

So far, we have considered classification of the Markov process based on the final state of the dynamics $z(\tau)$.
It is natural to consider a scenario in which classification is performed using the trajectory of the Markov process. 
The use of trajectory information for classification and estimation is often considered in modeling biochemical receptors \cite{Endres:2009:MLE,Mora:2010:MLE,Harvey:2023:energy-accuracy-tradeoffs,Pagare:2024:MpembaLikeSensor}. 
Let $\Gamma$ be a trajectory of a Markov process within the range $[0,\tau]$. 
Suppose there are $J$ jumps in the interval $[0,\tau]$. Let $t_j$ denote the time of the $j$-th jump, and let $z_j\in\mathfrak{B}$ be the state immediately after that jump. The trajectory $\Gamma$ can be written as
\begin{align}
    \Gamma=[(t_{0},z_{0}),(t_{1},z_{1}),\ldots,(t_{J},z_{J})],
    \label{eq:trajectory_single_def}
\end{align}
where $t_0=0$ is the initial time and $z_0$ is the initial state. 
Let $N_\circ(\Gamma)$ be a function of the trajectory, where $N_\circ(\Gamma)$ is constant for trajectories with no jumps.
Let $\Gamma_\varnothing$ be a trajectory with no jumps within the interval $[0,\tau]$.
$\Gamma_\varnothing$ can be expressed as $\Gamma_\varnothing=[(t_0,z_0)]$, 
containing only the initial state.
The condition for $N_\circ$ can be expressed as 
$N_\circ(\Gamma_\varnothing) = \mathrm{const}$. 
For instance, the counting observable is a particular example of $N_\circ(\Gamma)$. 
We consider a scenario in which the mapping function $T(\cdot)$ outputs the prediction through the trajectory function $N_\circ(\Gamma)$, i.e., $T(\Gamma)=T(N_\circ(\Gamma))$. 
For the case of trajectory-based classification, the initial states for $y=0$ and $y=1$ can be arbitrary.

We have derived Eqs.~\eqref{eq:result2} and \eqref{eq:result_da1} for classification based on the final state $z(\tau)$. 
Similarly, 
for trajectory-based classification, we can obtain lower bounds on the average error. 
Equations~\eqref{eq:result2} and \eqref{eq:result_da1} also hold for the trajectory-based classification scenario.
Specifically, when the final prediction $T(\cdot)$ is made via $N_\circ(\Gamma)$ starting from an arbitrary initial condition, Eqs.~\eqref{eq:result2} and \eqref{eq:result_da1} hold. 
We note that in state-based classification, Eq.~\eqref{eq:result1}, which involves entropy production, holds.
In contrast, we cannot obtain such a bound for trajectory-based classification. 
We summarize the results in Table I in the Supplementary Material \cite{Supp:2026:ThermoClassifier}. 
\nocite{Sason:2016:DivIneqReview,Pedregosa:2011:scikit-learn}
Moreover, we perform numerical simulations for the obtained bounds [Eqs.~\eqref{eq:result1}, \eqref{eq:result2}, and \eqref{eq:result_da1}], which are demonstrated in the Supplementary Material.

So far, we have considered the scenario in which the data are $y=0$ or $y=1$. 
Next, we consider the problem of classifying multiple data points.
Let $v$ be a variable representing a single data point. $v$ may be any type of data, such as a numerical value, a vector, or an image.
Here, $M$ data points are represented by $\mathcal{V}=\{v_1,v_2,\cdots,v_M\}$.
The goal is to classify each $v_i$ as $0$ or $1$ via a stochastic classifier, 
and each data point has a true class, denoted by $L(v_i)$ (ground truth). Here, the classes are assumed to be binary; that is, $L(v_i)\in \{0,1\}$.
We can divide the data points into $\mathcal{V}_0$ and $\mathcal{V}_1$, where $\mathcal{V}_c \equiv \{v \in \mathcal{V} \mid L(v)=c\}$. 
When we input a data point $v$ to the stochastic machine, we obtain an output $x$, where the conditional probability is $P(x\mid v)$.
Again, we apply the mapping function $T(x)$ to reduce the output $x$ to binary data, $T(x)\in \{0,1\}$.
We regard $T(x)$ as a prediction of the class label of the data point $v$. 
See Fig.~\ref{fig:classification_scenario}(b) for an illustrative example of this scenario. 
Given a data point $v$, the classification error in Eq.~\eqref{eq:Perr_y_def} becomes
\begin{align}
    P_{\mathrm{err}}(v)&=\sum_{x}\mathds{1}\{T(x)\ne L(v)\}P(x\mid v),
    \label{eq:Perr_v_def}
\end{align}
where $\mathds{1}\{T(x) \ne L(v)\}$ compares the true class $L(v)$ of $v$ with its prediction $T(x)$ by the classifier. 
In the previous scenario, we considered the average error in Eq.~\eqref{eq:Perr_Bayes_error}.
However, in this second scenario, it is difficult to derive a lower bound on the average error.
Here, we consider the worst-case error as follows:
\begin{align}
P^{\mathrm{wst}}_{\mathrm{err}}=\max_{v\in\mathcal{V}}P_{\mathrm{err}}(v),
    \label{eq:P_wrst_def}
\end{align}
which is the error probability of the most ``difficult'' data point in $\mathcal{V}$.
This kind of worst-case error is often considered in statistical decision theory \cite{Lehmann:1998:PointEstimation}.
According to Le Cam's two-point method \cite{LeCam:1973:dimensionality,Raginsky:2013:SLT,Wainwright:2019:HighDimensionalStatistics}, the lower bound on the worst-case error probability is 
\begin{align}
    P^{\mathrm{wst}}_{\mathrm{err}}\ge\frac{1}{2}\left[1-\mathrm{TV}\left(P(x\mid v),P(x\mid v^{\prime})\right)\right],
    \label{eq:Pwrt_err_lowerbound}
\end{align}
where $v$ and $v^\prime$ in Eq.~\eqref{eq:Pwrt_err_lowerbound} should be selected from $v \in \mathcal{V}_0$ and $v^\prime \in \mathcal{V}_1$. 
The optimal lower bound can be found by maximizing the right-hand side of Eq.~\eqref{eq:Pwrt_err_lowerbound} with respect to $v \in \mathcal{V}_0$ and $v^\prime \in \mathcal{V}_1$.

We now present an explicit classification process for the data points $\mathcal{V}$ based on the Markov process. 
Because we want to model $P(x\mid v)$ as a Markov process, we assume that the transition rate $W_{\nu\mu}$ depends on the data point $v$.
For the data point $v$, the transition rate is expressed as $W_{\nu\mu}^{[v]}$.
The probability $P(\mu;t)$ and state $z(t)$ corresponding to the Markov process of the data point $v$ are denoted by $P^{[v]}(\mu;t)$ and $z^{[v]}(t)$, respectively.
We assume that the initial state of the process is identical for all $v \in \mathcal{V}$. 
Here, we derive the lower bound on the worst-case error $P_\mathrm{err}^\mathrm{wst}$ in terms of the entropy production and dynamical activity of the classifier Markov process.
We obtain the lower bound on $P_\mathrm{err}^\mathrm{wst}$, which is identical to Eqs.~\eqref{eq:result1}, \eqref{eq:result2}, and \eqref{eq:result_da1}, where $\Sigma^{(0)}(t)+\Sigma^{(1)}(t)$, $\mathcal{A}^{(0)}(t)+\mathcal{A}^{(1)}(t)$, and $\mathfrak{a}^{(0)}(0)+\mathfrak{a}^{(1)}(0)$ are replaced with $\Sigma^{[v]}(t)+\Sigma^{[v^{\prime}]}(t)$, $\mathcal{A}^{[v]}(t)+\mathcal{A}^{[v^\prime]}(t)$, and $\mathfrak{a}^{[v]}(0)+\mathfrak{a}^{[v^\prime]}(0)$, respectively. 
As in Eq.~\eqref{eq:Pwrt_err_lowerbound}, $v$ and $v^\prime$ should be selected from $v \in \mathcal{V}_0$ and $v^\prime \in \mathcal{V}_1$.

We have considered data-point classification based on the final state $z(\tau)$. 
Again, we can consider classification of the data points based on a trajectory $\Gamma$. 
Similarly, 
for trajectory-based classification, we can obtain lower bounds on the worst-case error. 
Equations~\eqref{eq:result2} and \eqref{eq:result_da1} also hold for the trajectory-based classification scenario.
Specifically, when we use $N_\circ(\Gamma)$ for the mapping function $T(\cdot)$, Eqs.~\eqref{eq:result2} and \eqref{eq:result_da1} hold, where $\mathcal{A}^{(0)}(t)+\mathcal{A}^{(1)}(t)$ and $\mathfrak{a}^{(0)}(0)+\mathfrak{a}^{(1)}(0)$ are replaced with $\mathcal{A}^{[v]}(t)+\mathcal{A}^{[v^\prime]}(t)$ and $\mathfrak{a}^{[v]}(0)+\mathfrak{a}^{[v^\prime]}(0)$, respectively. 
As in Eq.~\eqref{eq:Pwrt_err_lowerbound}, $v$ and $v^\prime$ should be selected from $v \in \mathcal{V}_0$ and $v^\prime \in \mathcal{V}_1$. 
We summarize the results in Table II in the Supplementary Material \cite{Supp:2026:ThermoClassifier}.

Again, when $\Sigma^{[v]}(\tau)+\Sigma^{[v^\prime]}(\tau)$, $\mathcal{A}^{[v]}(\tau)+\mathcal{A}^{[v^\prime]}(\tau)$, or $\mathfrak{a}^{[v]}(0)+\mathfrak{a}^{[v^\prime]}(0)$ vanishes, the lower bound on the worst-case error becomes $1/2$, equivalent to random guessing. 
Although the bounds in the second scenario are similar to those in the first scenario, there is a degree of freedom in how to choose $v\in\mathcal{V}_0$ and $v^\prime \in \mathcal{V}_1$ in the derived bounds. 
Although the bounds themselves hold for arbitrary $v\in\mathcal{V}_0$ and $v^\prime \in \mathcal{V}_1$, the optimal bounds can be obtained by maximizing the lower bounds with respect to $v$ and $v^\prime$.

\textit{Discussion.---}
A key point in deriving Eqs.~\eqref{eq:result1}--\eqref{eq:result_da1} (and their counterparts in the classification of data points $\mathcal{V}$) in this study is to express the upper bound on the total variation distance in Eqs.~\eqref{eq:Bayes_error_classical} and \eqref{eq:Pwrt_err_lowerbound} in terms of physical quantities, such as entropy production or dynamical activity. 
In statistics, divergences (and the Fisher information)  
play a central role in deriving bounds.  
However, in general, it is difficult to express a divergence between arbitrary distributions in terms of physical quantities. 
Classification has been discussed in stochastic thermodynamics in terms of arrow-of-time inference \cite{Jarzynski:2011:Irreversibility,Roldan:2015:ArrowTimeDecision,Seif:2021:ThermoArrowTime}. 
However, this strategy only works for the discrimination of a specific pair of processes. 
This is because the Kullback-Leibler divergence between the forward and backward dynamics distributions corresponds to entropy production, but no such relationship exists between general distributions.
Cellular concentration sensing is a field related to the classification problem, which was pioneered by Ref.~\cite{Berg:1977:PhysicsOfChemoreception}. Several studies \cite{Endres:2009:MLE,Mora:2010:MLE,Harvey:2023:energy-accuracy-tradeoffs} assumed linear dependence on the ligand concentration, which allows the Fisher information to be reduced to dynamical activity or entropy production.
Other studies \cite{Berg:1977:PhysicsOfChemoreception,Mehta:2012:Energetic,Govern:2014:EnergyDissipationNoise} do not rely on the Fisher information, but the precision bounds were derived assuming specific models. 
In biochemical classification, several studies \cite{McKeithan:1995:KineticProofreading,Murugan:2014:DiscriminatoryProofreading,Hartich:2015:NonequilibriumSensing,Rao:2015:ThermodynamicsAccuracyProofreading,Ganti:2020:TCellSignaling,Berx:2024:tradeoffs-thermodynamics-energy-relay} used the pioneering Hopfield-Ninio model \cite{Hopfield:1974:KineticProofreading,Ninio:1975:KineticAmplification}, which assumes chain-like biochemical reactions. 
Reference~\cite{Wong:2018:EnergySpeedAccuracy} studied a generalization of the Hopfield-Ninio model, in which reaction networks are not restricted to a linear chain structure. Still, it is based on dissociation-based discrimination, and the reaction rates require several assumptions.  
In general classification problems, there is no consensus on how transition rates vary across classes.
Our derived bounds are general in that they hold regardless of how transition rates are assumed to be affected in each class.

This Letter derives an upper bound on the total variation distance through a derivation inspired by quantum information \cite{Peres:1991:OptimalDetection,Massar:1995:OptimalExtraction,Gill:2000:StateEstimation,Ekert:2002:DirectEstimations,Acin:2005:MultipleCopyDiscrimination}, but a derivation via a speed limit is also possible.
However, the resulting bounds via the speed limit are weaker, and the bound in Eq.~\eqref{eq:result_da1} cannot be derived from this alternative route (see Supplementary Material \cite{Supp:2026:ThermoClassifier}).

\textit{Conclusion.---}In this study, we have established a fundamental connection between the errors of binary classifiers and thermodynamic costs in Markovian dynamics. Specifically, we derived lower bounds on the average error and the worst-case error in terms of entropy production and dynamical activity for classifiers based on continuous-time Markov processes, considering both final-state and trajectory-based classification scenarios. 
Several directions remain open for future work. Exploring the extension of our bounds to specific machine learning architectures, such as physical neural network classifiers, may shed new light on the energetic limits of artificial intelligence.

\begin{acknowledgements}
This work was supported by JSPS KAKENHI Grant Numbers JP24K03008 and JP26K02998.
    
ChatGPT 5.4-5.6 was used to assist with code generation and checking for errors.
All scientific ideas, analyses, calculations, and conclusions are those of the author.
\end{acknowledgements}

\appendix

\onecolumngrid
\section*{End Matter}
\twocolumngrid
\dummyblock{
\section{Dummy}
}
\setcounter{equation}{0}
\renewcommand{\theequation}{A\arabic{equation}}

\EMsection{Combined Markov process} The derivation of the results [Eqs.~\eqref{eq:result1}, \eqref{eq:result2}, and \eqref{eq:result_da1}] is based on considering two Markov processes with \textit{no} interaction. 
We assume that the two Markov processes are identical except that the first process has the transition rate $W^{(0)}$, while the second process has the transition rate $W^{(1)}$. 
The state space of the combined process is the product of the two individual state sets.
Thus, the state set for the combined process has $|\mathfrak{B}^\oplus| = D^2$ states. 
Hereafter, a superscript $\oplus$ refers to a quantity of the combined system, whereas quantities with the superscript $(y)$ correspond to those of the class label $y$. 

Let $P^{\oplus}(\mu^{(0)},\mu^{(1)};t)$ denote the probability that the combined system is in state $(\mu^{(0)},\mu^{(1)})$ at time $t$.
We can identify the uncoupled two Markov processes as a single larger Markov process. 
The master equation for the combined process is
\begin{align}
    &\frac{d}{dt}P^{\oplus}(\nu^{(0)},\nu^{(1)};t)=\sum_{\mu^{(0)}}W^{(0)}_{\nu^{(0)}\mu^{(0)}}P^{\oplus}(\mu^{(0)},\nu^{(1)};t)\nonumber\\&+\sum_{\mu^{(1)}}W^{(1)}_{\nu^{(1)}\mu^{(1)}}P^{\oplus}(\nu^{(0)},\mu^{(1)};t).
    \label{eq:master_eq_two}
\end{align}
As in the single-process case [Eq.~\eqref{eq:master_equation}], we can describe the trajectories of the combined process.
Suppose the combined trajectory has $J$ jump events in the time interval $[0,\tau]$.
Let $t_j^{\oplus}$ be the time of the $j$-th jump, and let $z^{\oplus}_{j}\in\mathfrak{B}^{\oplus}$ be the post-jump state.
The trajectory is then written as
\begin{align}
    \Gamma^{\oplus}=[(t^{\oplus}_{0},z^{\oplus}_{0}),(t^{\oplus}_{1},z^{\oplus}_{1}),\ldots,(t^{\oplus}_{J},z^{\oplus}_{J})],
    \label{eq:trajectory_two}
\end{align}
where $t_0^{\oplus}=0$ and $z_{0}^{\oplus}$ is the initial state. 
For example, if the trajectories of the $y=0$ and $y=1$ processes are
\begin{align}
    \Gamma^{(0)}&=[(0,z^{(0)}_{0}),(1,z^{(0)}_{1})],\label{eq:Gamma_one}\\\Gamma^{(1)}&=[(0,z^{(1)}_{0}),(2,z^{(1)}_{1}),(3,z^{(1)}_{2})],\label{eq:Gamma_two}
\end{align}
then the combined trajectory is
\begin{align}
    \Gamma^{\oplus}&=[(0,(z^{(0)}_{0},z^{(1)}_{0})),(1,(z^{(0)}_{1},z^{(1)}_{0})),\nonumber\\&(2,(z^{(0)}_{1},z^{(1)}_{1})),(3,(z^{(0)}_{1},z^{(1)}_{2}))].
    \label{eq:Gamma_combined}
\end{align}
Because the two Markov processes are independent, the combined entropy production and dynamical activity are just the sums of the corresponding quantities for each process.

\EMsection{Derivation of the lower bound on the Bayes error}
We provide a sketch of the proof for the derivation of the result presented in Eq.~\eqref{eq:result1}. 
For details, please see the Supplementary Material \cite{Supp:2026:ThermoClassifier}. 

Using the inequality between the total variation distance and the Bhattacharyya coefficient, the Bayes error has the following lower bound \cite{Devroye:1996:pattern-recognition,Nielsen:2014:Generalized} (Chapter 3 of Ref.~\cite{Devroye:1996:pattern-recognition}):
\begin{align}
    P^{\mathrm{Bayes}}_{\mathrm{err}}\ge\frac{1}{2}-\frac{1}{2}\sqrt{1-4\pi_{0}\pi_{1}\mathrm{Bhat}\left(P(x\mid0),P(x\mid1)\right)^{2}},
    \label{eq:Bayes_Batt_lowerbound}
\end{align}
where $\mathrm{Bhat}(\cdot,\cdot)$ is the Bhattacharyya coefficient. 
We use the inequality for the $p$-norm of a vector to obtain a lower bound from Eq.~\eqref{eq:Bayes_Batt_lowerbound}. 
For an $n$-dimensional vector $u$, the $p$-norm is defined by $\|u\|_p=\left(\sum_{k=1}^n\left|u_k\right|^p\right)^{1 / p}$. For $0<p<q$, it is known that $\|u\|_p \geq\|u\|_q$. 
Using this relation in Eq.~\eqref{eq:Bayes_Batt_lowerbound}, the Bayes error is bounded from below by
\begin{align}
    P^{\mathrm{Bayes}}_{\mathrm{err}}\ge\frac{1}{2}-\frac{1}{2}\sqrt{1-4\pi_{0}\pi_{1}\sum_{x}P(x\mid0)P(x\mid1)}.
    \label{eq:Perr_min_lowerbound_classical}
\end{align}

Since the state at time $t=\tau$ is used for classification, $\sum_{x}P(x\mid0)P(x\mid1)$ in Eq.~\eqref{eq:Perr_min_lowerbound_classical} should be replaced by $\sum_{\mu}P^{(0)}(\mu;\tau)P^{(1)}(\mu;\tau)$
when considering the scenario of classification based on the final state $z(\tau)$. 
Here, $P^{(y)}(\mu;\tau)$ is the probability of the Markov process at time $t=\tau$ for the class label $y$. 
We wish to find a lower bound for $\sum_{\mu}P^{(0)}(\mu;\tau)P^{(1)}(\mu;\tau)$.
Let $O(\mu)$ be an observable of state $\mu$ of a Markov process. 
We define the expectation of $O(\mu)$ as follows:
\begin{align}
   \mathbb{E}_{P(\mu;t)}[O]\equiv\sum_{\mu}O(\mu)P(\mu;t).
   \label{eq:O_expectation}
\end{align}
Then, according to the Cram\'er-Rao inequality, the following relation holds:
\begin{align}
    \frac{\mathrm{Var}_{P(\nu;t)}[O]}{\left(\partial_{t}\mathbb{E}_{P(\nu;t)}[O]\right)^{2}}\ge\frac{1}{\mathcal{I}(t)},
    \label{eq:CrameRao}
\end{align}
where $\mathrm{Var}_{P(\mu;t)}[O]$ is the variance and $\mathcal{I}(t)$ is the temporal Fisher information.
The temporal Fisher information
quantifies the infinitesimal distance in probability space \cite{Wootters:1981:StatDist} and
plays an important role in thermodynamic trade-off relations \cite{Ito:2018:InfoGeo,Nicholson:2020:TIUncRel}.
It is known that the following relation holds \cite{Nishiyama:2026:TemporalFisherPRE}:
\begin{align}
    \mathcal{I}(t)\le\frac{\Sigma(t)}{2t^{2}},\,\,\,\mathcal{I}(t)\le\frac{\mathcal{A}(t)}{t^{2}},\label{eq:It_upperbound}
\end{align}
where $\Sigma(t)$ and $\mathcal{A}(t)$ are the time-integrated entropy production and time-integrated dynamical activity defined in Eqs.~\eqref{eq:EP_def} and \eqref{eq:integrated_activity_def}, respectively. 
Note that the entropy production bound in Eq.~\eqref{eq:It_upperbound} holds for Langevin dynamics, whereas the dynamical activity version does not.

Next, we consider the combined Markov process in the enlarged space, as explained above in Appendix~A. 
We consider two different Markov processes here. 
Let us consider the observable $O^{\oplus}(\mu^{(0)},\mu^{(1)})$, which is defined for the combined Markov process.
Since the combined Markov process comprises two independent Markov processes, $O^{\oplus}$ takes two arguments, corresponding to the states in the two Markov processes. 
The expectation of $O^\oplus$ is given by
\begin{align}
   &\mathbb{E}_{P^{\oplus}(\mu^{(0)},\mu^{(1)};\tau)}[O^{\oplus}]\nonumber\\&=\sum_{\mu^{(0)},\mu^{(1)}}O^{\oplus}(\mu^{(0)},\mu^{(1)})P^{\oplus}(\mu^{(0)},\mu^{(1)};\tau)\nonumber\\&=\sum_{\mu^{(0)},\mu^{(1)}}O^{\oplus}(\mu^{(0)},\mu^{(1)})P^{(0)}(\mu^{(0)};\tau)P^{(1)}(\mu^{(1)};\tau).
   \label{eq:O_expectation2}
\end{align}
Specifically, we employ the following observable:
\begin{align}
    \Omega^{\oplus}(\mu^{(0)},\mu^{(1)})=\begin{cases}
1 & \mu^{(0)}=\mu^{(1)}\\
0 & \mathrm{otherwise}
\end{cases}.
\label{eq:Oplus_def}
\end{align}
The observable expressed by Eq.~\eqref{eq:Oplus_def} is a quantity that returns $1$ if the final states of the two Markov processes match and $0$ otherwise.
This is reminiscent of the swap operator considered in the multi-copy measurement \cite{Peres:1991:OptimalDetection,Massar:1995:OptimalExtraction,Gill:2000:StateEstimation,Ekert:2002:DirectEstimations,Acin:2005:MultipleCopyDiscrimination}. 
We emphasize that the two Markov processes are not coupled; however, the observable defined in Eq.~\eqref{eq:Oplus_def} is coupled.
The expectation and variance of $\Omega^\oplus$ are given by
\begin{align}
    \mathbb{E}_{P^{\oplus}(\mu^{(0)},\mu^{(1)};\tau)}[\Omega^{\oplus}]&=\alpha(\tau),\label{eq:E_Omega}\\\mathrm{Var}_{P^{\oplus}(\mu^{(0)},\mu^{(1)};\tau)}[\Omega^{\oplus}]&=\alpha(\tau)-\alpha(\tau)^{2},\label{eq:Var_Omega}
\end{align}
where
\begin{align}
    \alpha(\tau)\equiv\sum_{\mu}P^{(0)}(\mu;\tau)P^{(1)}(\mu;\tau).
    \label{eq:alpha_tau_def}
\end{align} 
Substituting Eqs.~\eqref{eq:E_Omega} and \eqref{eq:Var_Omega} into Eq.~\eqref{eq:CrameRao}, we obtain
\begin{align}
    \alpha(\tau) \geq \cos\left(\frac{1}{2} \int_0^\tau \sqrt{\mathcal{I}^\oplus(t)} dt\right)^2,
    \label{eq:alphatau_bound}
\end{align}
which holds within the range $0\le\frac{1}{2}\int^{\tau}_{0}\sqrt{\mathcal{I}^\oplus(t)}dt\le\frac{\pi}{2}$. 
From Eq.~\eqref{eq:Perr_min_lowerbound_classical}, the Bayes error is
\begin{align}
    P^{\mathrm{Bayes}}_{\mathrm{err}}&\ge\frac{1}{2}-\frac{1}{2}\sqrt{1-4\pi_{0}\pi_{1}\alpha(\tau)}.
    \label{eq:Bayes_error_alphatau}
\end{align}
Substituting Eq.~\eqref{eq:alphatau_bound} into \eqref{eq:Bayes_error_alphatau}, we obtain
\begin{align}
    \min P^{\mathrm{Bayes}}_{\mathrm{err}}&\ge\frac{1}{2}-\frac{1}{2}\sqrt{1-4\pi_{0}\pi_{1}\cos\left(\frac{1}{2}\int^{\tau}_{0}\sqrt{\mathcal{I}^\oplus(t)}dt\right)^{2}}.
    \label{eq:Perr_min_It_bound}
\end{align}
Since $(\cos x)^2$ is a decreasing function for $0\le x \le \pi /2$, by using Eq.~\eqref{eq:It_upperbound} with the uniform prior $\pi_0=\pi_1=1/2$, we have Eqs.~\eqref{eq:result1} and \eqref{eq:result2} in the main text. 

The inequality in Eq.~\eqref{eq:result2} also holds for trajectory-based classification. 
In the combined Markov process, 
the relation in Ref.~\cite{Terlizzi:2019:KUR} holds:
\begin{align}
    \frac{\mathrm{Var}\left[N^{\oplus}\right]}{\tau^{2}\left(\partial_{\tau}\mathbb{E}\left[N^{\oplus}\right]\right)^{2}}\geq\frac{1}{\mathcal{A}^{\oplus}(\tau)}.
    \label{eq:KUR}
\end{align}
Here, $N^\oplus(\Gamma)$ is an arbitrary trajectory observable defined in the combined Markov process, and the expectation and variance are evaluated with respect to the path probability. 
We repeat the same procedure for Eq.~\eqref{eq:KUR} to derive Eq.~\eqref{eq:result2} for the trajectory classification scenario (see the Supplementary Material \cite{Supp:2026:ThermoClassifier}).

\EMsection{Derivation of an alternative lower bound on the Bayes error}
We derive Eq.~\eqref{eq:result_da1} in the main text using the uncertainty relation from Ref.~\cite{Hasegawa:2024:ConcentrationIneqPRL} rather than the Cram\'er-Rao inequality.
First, we consider a Markov process expressed by Eq.~\eqref{eq:master_equation}.
Let us consider the observable $N_{\varnothing}$, which vanishes for $\Gamma_\varnothing$, $N_\varnothing(\Gamma_\varnothing)=0$. 
Then, the following relation holds \cite{Hasegawa:2024:ConcentrationIneqPRL}:
\begin{align}
    \frac{\mathrm{Var}[N_{\varnothing}]}{\mathbb{E}[N_{\varnothing}]^{2}}\ge\frac{1}{e^{\mathfrak{a}(0)\tau}-1},
    \label{eq:concentration_transient_KUR}
\end{align}
where $\mathfrak{a}(0)$ is the dynamical activity at time $t=0$.
Consider Eq.~\eqref{eq:concentration_transient_KUR} in the combined Markov process.
Then we have
\begin{align}
    \frac{\mathrm{Var}[N^{\oplus}_{\varnothing}]}{\mathbb{E}[N^{\oplus}_{\varnothing}]^{2}}&\ge\frac{1}{e^{\mathfrak{a}^{\oplus}(0)\tau}-1}.
    \label{eq:concentration_transient_KUR_combined}
\end{align}
Here, $N_\varnothing^\oplus$ is the observable in the combined Markov process that vanishes for no-jump trajectories.

Similar to the observable considered in Eq.~\eqref{eq:Oplus_def}, we employ the following observable for $N_\varnothing^\oplus$:
\begin{align}
    \Phi^{\oplus}_{\varnothing}(\Gamma^{\oplus})\equiv\begin{cases}
0 & z^{(0)}(\tau)=z^{(1)}(\tau)\\
1 & \mathrm{otherwise}.
\end{cases}
\label{eq:Phi_oplus_def}
\end{align}
In the main text, we have assumed that the initial state is identical for $y=0$ and $y=1$, i.e., $z^{(0)}(0)=z^{(1)}(0)$. Consequently, if no jump occurs in $\Gamma^\oplus$, then $z^{(0)}(\tau)=z^{(1)}(\tau)$, and $\Phi^{\oplus}_{\varnothing}(\Gamma^{\oplus}) = 0$ for $\Gamma_\varnothing^\oplus$. Equation~\eqref{eq:concentration_transient_KUR_combined} therefore holds for $\Phi^{\oplus}_{\varnothing}$, and thus $\Phi_\varnothing^\oplus$ can be used in Eq.~\eqref{eq:concentration_transient_KUR_combined} in place of $N_\varnothing^\oplus$.
The expectation and variance for Eq.~\eqref{eq:Phi_oplus_def} are given by
\begin{align}
    \mathbb{E}\left[\Phi^{\oplus}_{\varnothing}(\Gamma^{\oplus})\right]&=1-\alpha(\tau),\label{eq:Phi_expectation}\\\mathrm{Var}[\Phi^{\oplus}_{\varnothing}(\Gamma^{\oplus})]&=1-\alpha(\tau)-(1-\alpha(\tau))^{2},\label{eq:Phi_variance}
\end{align}
where $\alpha(\tau)$ is defined in Eq.~\eqref{eq:alpha_tau_def}. 
Substituting Eqs.~\eqref{eq:Phi_expectation} and \eqref{eq:Phi_variance} into Eq.~\eqref{eq:concentration_transient_KUR_combined}, we obtain
\begin{align}
    e^{-\mathfrak{a}^{\oplus}(0)\tau}\le\alpha(\tau).
    \label{eq:alphatau_lowerbound}
\end{align}
Substituting Eq.~\eqref{eq:alphatau_lowerbound} into Eq.~\eqref{eq:Bayes_error_alphatau} and using $\pi_0=\pi_1=1/2$,
we obtain Eq.~\eqref{eq:result_da1} in the main text. 

Equation~\eqref{eq:result_da1} holds for trajectory-based classification as well. 
Please see the Supplementary Material for details \cite{Supp:2026:ThermoClassifier}.

\EMsection{Derivation based on Le Cam's two-point method}
We derive Eq.~\eqref{eq:Pwrt_err_lowerbound} in the main text. 
The derivation follows Le Cam's well-known two-point method \cite{LeCam:1973:dimensionality,Raginsky:2013:SLT,Wainwright:2019:HighDimensionalStatistics}.
The worst-case classification error
$P^{\mathrm{wst}}_{\mathrm{err}}=\max_{v\in\mathcal{V}}\sum_{x}\mathds{1}\{T(x)\ne L(v)\}P(x\mid v)$ has the following lower bound:
\begin{align}
    &\max_{v\in\mathcal{V}}\sum_{x}\mathds{1}\{T(x)\ne L(v)\}P(x\mid v)\nonumber\\&\ge\max_{v\in\{v_{0},v_{1}\}}\sum_{x}\mathds{1}\{T(x)\ne L(v)\}P(x\mid v),
    \label{eq:LeCam_twopoint_before}
\end{align}
where $v_0$ and $v_1$ should be chosen from $\mathcal{V}_0$ and $\mathcal{V}_1$, respectively. 
The last term in Eq.~\eqref{eq:LeCam_twopoint_before} has the following lower bound:
\begin{align}
&\max_{v\in\{v_{0},v_{1}\}}\sum_{x}\mathds{1}\{T(x)\ne L(v)\}P(x\mid v)\nonumber\\&\ge\frac{1}{2}\sum_{x}\bigl[\mathds{1}\{T(x)\ne L(v_{0})\}P(x\mid v_{0})\bigr)\nonumber\\&+\mathds{1}\{T(x)\ne L(v_{1})\}P(x\mid v_{1})\bigr]\nonumber\\&\ge\sum_{x}\frac{\mathds{1}\{T(x)\ne L(v_{0})\}+\mathds{1}\{T(x)\ne L(v_{1})\}}{2}\nonumber\\&\times\min\left(P(x\mid v_{0}),P(x\mid v_{1})\right)\nonumber\\&\ge\frac{1}{2}\sum_{x}\min\left(P(x\mid v_{0}),P(x\mid v_{1})\right)\nonumber\\&=\frac{1}{2}\left(1-\mathrm{TV}\left(P(x\mid v_{0}),P(x\mid v_{1})\right)\right).
\label{eq:LeCam_derivation}
\end{align}
From the fourth to the sixth line, we used the property $\mathds{1}\{T(x)\ne L(v_{0})\}+\mathds{1}\{T(x)\ne L(v_{1})\}=1$ because $v_0\in\mathcal{V}_0$ and $v_1 \in \mathcal{V}_1$. 
From the sixth to the seventh line, $\min(a,b)=(a+b)/2-|b-a|/2$ is used. 
In summary, the worst-case error is bounded from below by 
\begin{align}
    P^{\mathrm{wst}}_{\mathrm{err}}\ge\frac{1}{2}\left(1-\mathrm{TV}\left(P(x\mid v),P(x\mid v^\prime)\right)\right),
    \label{eq:Perr_wrt_TV_lowerbound}
\end{align}
where $v \in \mathcal{V}_0$ and $v^\prime \in \mathcal{V}_1$. 
Equation~\eqref{eq:Perr_wrt_TV_lowerbound} is Eq.~\eqref{eq:Pwrt_err_lowerbound} in the main text.
Since Eq.~\eqref{eq:Perr_wrt_TV_lowerbound} is identical to Eq.~\eqref{eq:Bayes_error_classical} with $\pi_0=\pi_1=1/2$, subsequent calculations are essentially the same. 
While Eq.~\eqref{eq:Bayes_error_classical} concerns $P(x\mid y=0)$ and $P(x\mid y=1)$, Eq.~\eqref{eq:Perr_wrt_TV_lowerbound} concerns $P(x\mid v)$ and $P(x\mid v^\prime)$ ($v \in \mathcal{V}_0$ and $v^\prime \in \mathcal{V}_1$). 
This indicates that $\Sigma^{(0)}(t)+\Sigma^{(1)}(t)$, $\mathcal{A}^{(0)}(t)+\mathcal{A}^{(1)}(t)$, and $\mathfrak{a}^{(0)}(0)+\mathfrak{a}^{(1)}(0)$ in Eqs.~\eqref{eq:result1}, \eqref{eq:result2}, and \eqref{eq:result_da1} should be replaced with $\Sigma^{[v]}(t)+\Sigma^{[v^\prime]}(t)$, $\mathcal{A}^{[v]}(t)+\mathcal{A}^{[v^\prime]}(t)$, and $\mathfrak{a}^{[v]}(0)+\mathfrak{a}^{[v^\prime]}(0)$, respectively. 
Then, we obtain the following lower bounds for classification based on the final state $z(\tau)$:
\begin{align}
    P^{\mathrm{wst}}_{\mathrm{err}}&\geq\frac{1}{2}\left[1-\sin\left(\frac{\int^{\tau}_{0}\frac{\sqrt{\Sigma^{[v]}(t)+\Sigma^{[v^{\prime}]}(t)}}{t}dt}{2\sqrt{2}}\right)\right],
    \label{eq:result1_wst}\\
    P^{\mathrm{wst}}_{\mathrm{err}}&\ge\frac{1}{2}\left[1-\sin\left(\frac{\int^{\tau}_{0}\frac{\sqrt{\mathcal{A}^{[v]}(t)+\mathcal{A}^{[v^{\prime}]}(t)}}{t}dt}{2}\right)\right],
    \label{eq:result2_wst}\\
    P^{\mathrm{wst}}_{\mathrm{err}}&\ge\frac{1}{2}\left(1-\sqrt{1-e^{-\tau\left[\mathfrak{a}^{[v]}(0)+\mathfrak{a}^{[v^{\prime}]}(0)\right]}}\right),
    \label{eq:result_da1_wst}
\end{align}
where $v \in \mathcal{V}_0$ and $v^\prime \in \mathcal{V}_1$. 
For classification based on trajectory information, Eqs.~\eqref{eq:result2_wst} and \eqref{eq:result_da1_wst} hold.


\begin{thebibliography}{68}%
\makeatletter
\providecommand \@ifxundefined [1]{%
 \@ifx{#1\undefined}
}%
\providecommand \@ifnum [1]{%
 \ifnum #1\expandafter \@firstoftwo
 \else \expandafter \@secondoftwo
 \fi
}%
\providecommand \@ifx [1]{%
 \ifx #1\expandafter \@firstoftwo
 \else \expandafter \@secondoftwo
 \fi
}%
\providecommand \natexlab [1]{#1}%
\providecommand \enquote  [1]{``#1''}%
\providecommand \bibnamefont  [1]{#1}%
\providecommand \bibfnamefont [1]{#1}%
\providecommand \citenamefont [1]{#1}%
\providecommand \href@noop [0]{\@secondoftwo}%
\providecommand \href [0]{\begingroup \@sanitize@url \@href}%
\providecommand \@href[1]{\@@startlink{#1}\@@href}%
\providecommand \@@href[1]{\endgroup#1\@@endlink}%
\providecommand \@sanitize@url [0]{\catcode `\\12\catcode `\$12\catcode `\&12\catcode `\#12\catcode `\^12\catcode `\_12\catcode `\%12\relax}%
\providecommand \@@startlink[1]{}%
\providecommand \@@endlink[0]{}%
\providecommand \url  [0]{\begingroup\@sanitize@url \@url }%
\providecommand \@url [1]{\endgroup\@href {#1}{\urlprefix }}%
\providecommand \urlprefix  [0]{URL }%
\providecommand \Eprint [0]{\href }%
\providecommand \doibase [0]{https://doi.org/}%
\providecommand \selectlanguage [0]{\@gobble}%
\providecommand \bibinfo  [0]{\@secondoftwo}%
\providecommand \bibfield  [0]{\@secondoftwo}%
\providecommand \translation [1]{[#1]}%
\providecommand \BibitemOpen [0]{}%
\providecommand \bibitemStop [0]{}%
\providecommand \bibitemNoStop [0]{.\EOS\space}%
\providecommand \EOS [0]{\spacefactor3000\relax}%
\providecommand \BibitemShut  [1]{\csname bibitem#1\endcsname}%
\let\auto@bib@innerbib\@empty
\bibitem [{\citenamefont {Landauer}(1961)}]{Landauer:1961:LP}%
  \BibitemOpen
  \bibfield  {author} {\bibinfo {author} {\bibfnamefont {R.}~\bibnamefont {Landauer}},\ }\bibfield  {title} {\bibinfo {title} {Irreversibility and heat generation in the computing process},\ }\href {https://doi.org/10.1147/rd.53.0183} {\bibfield  {journal} {\bibinfo  {journal} {IBM J. Res. Dev.}\ }\textbf {\bibinfo {volume} {5}},\ \bibinfo {pages} {183} (\bibinfo {year} {1961})}\BibitemShut {NoStop}%
\bibitem [{\citenamefont {Bennett}(2003)}]{Bennett:2003:Landauer}%
  \BibitemOpen
  \bibfield  {author} {\bibinfo {author} {\bibfnamefont {C.~H.}\ \bibnamefont {Bennett}},\ }\bibfield  {title} {\bibinfo {title} {Notes on {Landauer}'s principle, reversible computation, and {Maxwell}'s demon},\ }\href {https://doi.org/10.1016/S1355-2198(03)00039-X} {\bibfield  {journal} {\bibinfo  {journal} {Stud. Hist. Philos. Mod. Phys.}\ }\textbf {\bibinfo {volume} {34}},\ \bibinfo {pages} {501} (\bibinfo {year} {2003})}\BibitemShut {NoStop}%
\bibitem [{\citenamefont {Bennett}(1973)}]{Bennett:1973:LogicalReversibility}%
  \BibitemOpen
  \bibfield  {author} {\bibinfo {author} {\bibfnamefont {C.~H.}\ \bibnamefont {Bennett}},\ }\bibfield  {title} {\bibinfo {title} {Logical reversibility of computation},\ }\href {https://doi.org/10.1147/rd.176.0525} {\bibfield  {journal} {\bibinfo  {journal} {IBM J. Res. Dev.}\ }\textbf {\bibinfo {volume} {17}},\ \bibinfo {pages} {525} (\bibinfo {year} {1973})}\BibitemShut {NoStop}%
\bibitem [{\citenamefont {Zurek}(1989)}]{Zurek:1989:ThermodynamicCostComputation}%
  \BibitemOpen
  \bibfield  {author} {\bibinfo {author} {\bibfnamefont {W.~H.}\ \bibnamefont {Zurek}},\ }\bibfield  {title} {\bibinfo {title} {Thermodynamic cost of computation, algorithmic complexity and the information metric},\ }\href {https://doi.org/10.1038/341119a0} {\bibfield  {journal} {\bibinfo  {journal} {Nature}\ }\textbf {\bibinfo {volume} {341}},\ \bibinfo {pages} {119} (\bibinfo {year} {1989})}\BibitemShut {NoStop}%
\bibitem [{\citenamefont {Seifert}(2012)}]{Seifert:2012:FTReview}%
  \BibitemOpen
  \bibfield  {author} {\bibinfo {author} {\bibfnamefont {U.}~\bibnamefont {Seifert}},\ }\bibfield  {title} {\bibinfo {title} {Stochastic thermodynamics, fluctuation theorems and molecular machines},\ }\href {http://stacks.iop.org/0034-4885/75/i=12/a=126001} {\bibfield  {journal} {\bibinfo  {journal} {Rep. Prog. Phys.}\ }\textbf {\bibinfo {volume} {75}},\ \bibinfo {pages} {126001} (\bibinfo {year} {2012})}\BibitemShut {NoStop}%
\bibitem [{\citenamefont {Sagawa}(2013)}]{Sagawa:2013:InformationProcessing}%
  \BibitemOpen
  \bibfield  {author} {\bibinfo {author} {\bibfnamefont {T.}~\bibnamefont {Sagawa}},\ }\href {https://doi.org/10.1007/978-4-431-54168-4} {\emph {\bibinfo {title} {Thermodynamics of Information Processing in Small Systems}}},\ \bibinfo {edition} {1st}\ ed.,\ Springer Theses\ (\bibinfo {year} {2013})\ pp.\ \bibinfo {pages} {XVI, 120}\BibitemShut {NoStop}%
\bibitem [{\citenamefont {Kolchinsky}\ and\ \citenamefont {Wolpert}(2017)}]{Kolchinsky:2017:DissipationInitialDistribution}%
  \BibitemOpen
  \bibfield  {author} {\bibinfo {author} {\bibfnamefont {A.}~\bibnamefont {Kolchinsky}}\ and\ \bibinfo {author} {\bibfnamefont {D.~H.}\ \bibnamefont {Wolpert}},\ }\bibfield  {title} {\bibinfo {title} {Dependence of dissipation on the initial distribution over states},\ }\href {https://doi.org/10.1088/1742-5468/aa7ee1} {\bibfield  {journal} {\bibinfo  {journal} {J. Stat. Mech.}\ }\textbf {\bibinfo {volume} {2017}},\ \bibinfo {pages} {083202} (\bibinfo {year} {2017})}\BibitemShut {NoStop}%
\bibitem [{\citenamefont {Wolpert}\ and\ \citenamefont {Kolchinsky}(2020)}]{Wolpert:2020:ThermodynamicsComputingCircuits}%
  \BibitemOpen
  \bibfield  {author} {\bibinfo {author} {\bibfnamefont {D.~H.}\ \bibnamefont {Wolpert}}\ and\ \bibinfo {author} {\bibfnamefont {A.}~\bibnamefont {Kolchinsky}},\ }\bibfield  {title} {\bibinfo {title} {Thermodynamics of computing with circuits},\ }\href {https://doi.org/10.1088/1367-2630/ab82b8} {\bibfield  {journal} {\bibinfo  {journal} {New J. Phys.}\ }\textbf {\bibinfo {volume} {22}},\ \bibinfo {pages} {063047} (\bibinfo {year} {2020})}\BibitemShut {NoStop}%
\bibitem [{\citenamefont {Manzano}\ \emph {et~al.}(2024)\citenamefont {Manzano}, \citenamefont {Karde{\c{s}}}, \citenamefont {Rold{\'a}n},\ and\ \citenamefont {Wolpert}}]{Manzano:2024:AbsoluteIrreversibility}%
  \BibitemOpen
  \bibfield  {author} {\bibinfo {author} {\bibfnamefont {G.}~\bibnamefont {Manzano}}, \bibinfo {author} {\bibfnamefont {G.}~\bibnamefont {Karde{\c{s}}}}, \bibinfo {author} {\bibfnamefont {{\'E}.}~\bibnamefont {Rold{\'a}n}},\ and\ \bibinfo {author} {\bibfnamefont {D.~H.}\ \bibnamefont {Wolpert}},\ }\bibfield  {title} {\bibinfo {title} {Thermodynamics of computations with absolute irreversibility, unidirectional transitions, and stochastic computation times},\ }\href {https://doi.org/10.1103/PhysRevX.14.021026} {\bibfield  {journal} {\bibinfo  {journal} {Phys. Rev. X}\ }\textbf {\bibinfo {volume} {14}},\ \bibinfo {pages} {021026} (\bibinfo {year} {2024})}\BibitemShut {NoStop}%
\bibitem [{\citenamefont {Goldt}\ and\ \citenamefont {Seifert}(2017)}]{Goldt:2017:STLearning}%
  \BibitemOpen
  \bibfield  {author} {\bibinfo {author} {\bibfnamefont {S.}~\bibnamefont {Goldt}}\ and\ \bibinfo {author} {\bibfnamefont {U.}~\bibnamefont {Seifert}},\ }\bibfield  {title} {\bibinfo {title} {Stochastic thermodynamics of learning},\ }\href {https://doi.org/10.1103/PhysRevLett.118.010601} {\bibfield  {journal} {\bibinfo  {journal} {Phys. Rev. Lett.}\ }\textbf {\bibinfo {volume} {118}},\ \bibinfo {pages} {010601} (\bibinfo {year} {2017})}\BibitemShut {NoStop}%
\bibitem [{\citenamefont {Horowitz}\ and\ \citenamefont {Esposito}(2014)}]{Horowitz:2014:InfoFlow}%
  \BibitemOpen
  \bibfield  {author} {\bibinfo {author} {\bibfnamefont {J.~M.}\ \bibnamefont {Horowitz}}\ and\ \bibinfo {author} {\bibfnamefont {M.}~\bibnamefont {Esposito}},\ }\bibfield  {title} {\bibinfo {title} {Thermodynamics with continuous information flow},\ }\href {https://link.aps.org/doi/10.1103/PhysRevX.4.031015} {\bibfield  {journal} {\bibinfo  {journal} {Phys. Rev. X}\ }\textbf {\bibinfo {volume} {4}},\ \bibinfo {pages} {031015} (\bibinfo {year} {2014})}\BibitemShut {NoStop}%
\bibitem [{\citenamefont {Kolchinsky}\ and\ \citenamefont {Wolpert}(2020)}]{Kolchinsky:2020:ThermodynamicCosts}%
  \BibitemOpen
  \bibfield  {author} {\bibinfo {author} {\bibfnamefont {A.}~\bibnamefont {Kolchinsky}}\ and\ \bibinfo {author} {\bibfnamefont {D.~H.}\ \bibnamefont {Wolpert}},\ }\bibfield  {title} {\bibinfo {title} {Thermodynamic costs of turing machines},\ }\href {https://doi.org/10.1103/PhysRevResearch.2.033312} {\bibfield  {journal} {\bibinfo  {journal} {Phys. Rev. Res.}\ }\textbf {\bibinfo {volume} {2}},\ \bibinfo {pages} {033312} (\bibinfo {year} {2020})}\BibitemShut {NoStop}%
\bibitem [{\citenamefont {Ikeda}\ \emph {et~al.}(2025)\citenamefont {Ikeda}, \citenamefont {Uda}, \citenamefont {Okanohara},\ and\ \citenamefont {Ito}}]{Ikeda:2025:SpeedAccuracyDiffusion}%
  \BibitemOpen
  \bibfield  {author} {\bibinfo {author} {\bibfnamefont {K.}~\bibnamefont {Ikeda}}, \bibinfo {author} {\bibfnamefont {T.}~\bibnamefont {Uda}}, \bibinfo {author} {\bibfnamefont {D.}~\bibnamefont {Okanohara}},\ and\ \bibinfo {author} {\bibfnamefont {S.}~\bibnamefont {Ito}},\ }\bibfield  {title} {\bibinfo {title} {Speed-accuracy relations for diffusion models: Wisdom from nonequilibrium thermodynamics and optimal transport},\ }\href {https://doi.org/10.1103/x5vj-8jq9} {\bibfield  {journal} {\bibinfo  {journal} {Phys. Rev. X}\ }\textbf {\bibinfo {volume} {15}},\ \bibinfo {pages} {031031} (\bibinfo {year} {2025})}\BibitemShut {NoStop}%
\bibitem [{\citenamefont {Wolpert}(2019)}]{Wolpert:2019:StochasticComputation}%
  \BibitemOpen
  \bibfield  {author} {\bibinfo {author} {\bibfnamefont {D.~H.}\ \bibnamefont {Wolpert}},\ }\bibfield  {title} {\bibinfo {title} {The stochastic thermodynamics of computation},\ }\href {https://doi.org/10.1088/1751-8121/ab0850} {\bibfield  {journal} {\bibinfo  {journal} {J. Phys. A: Math. Theor.}\ }\textbf {\bibinfo {volume} {52}},\ \bibinfo {pages} {193001} (\bibinfo {year} {2019})}\BibitemShut {NoStop}%
\bibitem [{\citenamefont {Wolpert}\ \emph {et~al.}(2024)\citenamefont {Wolpert}, \citenamefont {Korbel}, \citenamefont {Lynn}, \citenamefont {Tasnim}, \citenamefont {Grochow}, \citenamefont {Karde{\c{s}}}, \citenamefont {Aimone}, \citenamefont {Balasubramanian}, \citenamefont {De~Giuli}, \citenamefont {Doty}, \citenamefont {Freitas}, \citenamefont {Marsili}, \citenamefont {Ouldridge}, \citenamefont {Richa}, \citenamefont {Riechers}, \citenamefont {Rold{\'a}n}, \citenamefont {Rubenstein}, \citenamefont {Toroczkai},\ and\ \citenamefont {Paradiso}}]{Wolpert:2024:stochastic-thermodynamics-computation}%
  \BibitemOpen
  \bibfield  {author} {\bibinfo {author} {\bibfnamefont {D.~H.}\ \bibnamefont {Wolpert}}, \bibinfo {author} {\bibfnamefont {J.}~\bibnamefont {Korbel}}, \bibinfo {author} {\bibfnamefont {C.~W.}\ \bibnamefont {Lynn}}, \bibinfo {author} {\bibfnamefont {F.}~\bibnamefont {Tasnim}}, \bibinfo {author} {\bibfnamefont {J.~A.}\ \bibnamefont {Grochow}}, \bibinfo {author} {\bibfnamefont {G.}~\bibnamefont {Karde{\c{s}}}}, \bibinfo {author} {\bibfnamefont {J.~B.}\ \bibnamefont {Aimone}}, \bibinfo {author} {\bibfnamefont {V.}~\bibnamefont {Balasubramanian}}, \bibinfo {author} {\bibfnamefont {E.}~\bibnamefont {De~Giuli}}, \bibinfo {author} {\bibfnamefont {D.}~\bibnamefont {Doty}}, \bibinfo {author} {\bibfnamefont {N.}~\bibnamefont {Freitas}}, \bibinfo {author} {\bibfnamefont {M.}~\bibnamefont {Marsili}}, \bibinfo {author} {\bibfnamefont {T.~E.}\ \bibnamefont {Ouldridge}}, \bibinfo {author} {\bibfnamefont {A.~W.}\ \bibnamefont {Richa}}, \bibinfo {author} {\bibfnamefont {P.}~\bibnamefont {Riechers}}, \bibinfo {author}
  {\bibfnamefont {{\'E}.}~\bibnamefont {Rold{\'a}n}}, \bibinfo {author} {\bibfnamefont {B.}~\bibnamefont {Rubenstein}}, \bibinfo {author} {\bibfnamefont {Z.}~\bibnamefont {Toroczkai}},\ and\ \bibinfo {author} {\bibfnamefont {J.}~\bibnamefont {Paradiso}},\ }\bibfield  {title} {\bibinfo {title} {Is stochastic thermodynamics the key to understanding the energy costs of computation?},\ }\href {https://doi.org/10.1073/pnas.2321112121} {\bibfield  {journal} {\bibinfo  {journal} {Proc. Natl. Acad. Sci. U.S.A.}\ }\textbf {\bibinfo {volume} {121}},\ \bibinfo {pages} {e2321112121} (\bibinfo {year} {2024})}\BibitemShut {NoStop}%
\bibitem [{\citenamefont {Rabiner}(1989)}]{Rabiner:1989:hmm-tutorial}%
  \BibitemOpen
  \bibfield  {author} {\bibinfo {author} {\bibfnamefont {L.~R.}\ \bibnamefont {Rabiner}},\ }\bibfield  {title} {\bibinfo {title} {A tutorial on hidden markov models and selected applications in speech recognition},\ }\href {https://doi.org/10.1109/5.18626} {\bibfield  {journal} {\bibinfo  {journal} {Proc. IEEE}\ }\textbf {\bibinfo {volume} {77}},\ \bibinfo {pages} {257} (\bibinfo {year} {1989})}\BibitemShut {NoStop}%
\bibitem [{\citenamefont {Smyth}(1996)}]{Smyth:1997:hmm-clustering}%
  \BibitemOpen
  \bibfield  {author} {\bibinfo {author} {\bibfnamefont {P.}~\bibnamefont {Smyth}},\ }\bibfield  {title} {\bibinfo {title} {Clustering sequences with hidden {Markov} models},\ }in\ \href {https://proceedings.neurips.cc/paper_files/paper/1996/file/6a61d423d02a1c56250dc23ae7ff12f3-Paper.pdf} {\emph {\bibinfo {booktitle} {Adv. Neural Inf. Process. Syst.}}},\ Vol.~\bibinfo {volume} {9},\ \bibinfo {editor} {edited by\ \bibinfo {editor} {\bibfnamefont {M.~C.}\ \bibnamefont {Mozer}}, \bibinfo {editor} {\bibfnamefont {M.}~\bibnamefont {Jordan}},\ and\ \bibinfo {editor} {\bibfnamefont {T.}~\bibnamefont {Petsche}}}\ (\bibinfo {year} {1996})\BibitemShut {NoStop}%
\bibitem [{\citenamefont {Bishop}(2006)}]{Bishop:2006:PRML}%
  \BibitemOpen
  \bibfield  {author} {\bibinfo {author} {\bibfnamefont {C.~M.}\ \bibnamefont {Bishop}},\ }\href@noop {} {\emph {\bibinfo {title} {Pattern Recognition and Machine Learning}}},\ Information Science and Statistics\ (\bibinfo  {publisher} {Springer},\ \bibinfo {address} {New York},\ \bibinfo {year} {2006})\BibitemShut {NoStop}%
\bibitem [{\citenamefont {Esling}\ and\ \citenamefont {Agon}(2012)}]{Esling:2012:TimeSeriesDataMining}%
  \BibitemOpen
  \bibfield  {author} {\bibinfo {author} {\bibfnamefont {P.}~\bibnamefont {Esling}}\ and\ \bibinfo {author} {\bibfnamefont {C.}~\bibnamefont {Agon}},\ }\bibfield  {title} {\bibinfo {title} {Time-series data mining},\ }\href {https://doi.org/10.1145/2379776.2379788} {\bibfield  {journal} {\bibinfo  {journal} {ACM Comput. Surv.}\ }\textbf {\bibinfo {volume} {45}} (\bibinfo {year} {2012})}\BibitemShut {NoStop}%
\bibitem [{\citenamefont {Panuccio}\ \emph {et~al.}(2002)\citenamefont {Panuccio}, \citenamefont {Bicego},\ and\ \citenamefont {Murino}}]{Panuccio:2002:sequential-clustering}%
  \BibitemOpen
  \bibfield  {author} {\bibinfo {author} {\bibfnamefont {A.}~\bibnamefont {Panuccio}}, \bibinfo {author} {\bibfnamefont {M.}~\bibnamefont {Bicego}},\ and\ \bibinfo {author} {\bibfnamefont {V.}~\bibnamefont {Murino}},\ }\bibfield  {title} {\bibinfo {title} {A hidden markov model-based approach to sequential data clustering},\ }in\ \href {https://doi.org/10.1007/3-540-70659-3_77} {\emph {\bibinfo {booktitle} {Struct., Syntactic, Stat. Pattern Recognit.}}},\ \bibinfo {editor} {edited by\ \bibinfo {editor} {\bibfnamefont {T.}~\bibnamefont {Caelli}}, \bibinfo {editor} {\bibfnamefont {A.}~\bibnamefont {Amin}}, \bibinfo {editor} {\bibfnamefont {R.~P.~W.}\ \bibnamefont {Duin}}, \bibinfo {editor} {\bibfnamefont {D.}~\bibnamefont {de~Ridder}},\ and\ \bibinfo {editor} {\bibfnamefont {M.}~\bibnamefont {Kamel}}}\ (\bibinfo {year} {2002})\ pp.\ \bibinfo {pages} {734--743}\BibitemShut {NoStop}%
\bibitem [{\citenamefont {Stella}\ and\ \citenamefont {Amer}(2012)}]{Stella:2012:ctbn-classifiers}%
  \BibitemOpen
  \bibfield  {author} {\bibinfo {author} {\bibfnamefont {F.}~\bibnamefont {Stella}}\ and\ \bibinfo {author} {\bibfnamefont {Y.}~\bibnamefont {Amer}},\ }\bibfield  {title} {\bibinfo {title} {Continuous time {B}ayesian network classifiers},\ }\href {https://doi.org/10.1016/j.jbi.2012.07.002} {\bibfield  {journal} {\bibinfo  {journal} {J. Biomed. Inform.}\ }\textbf {\bibinfo {volume} {45}},\ \bibinfo {pages} {1108} (\bibinfo {year} {2012})}\BibitemShut {NoStop}%
\bibitem [{\citenamefont {Spaeh}\ and\ \citenamefont {Tsourakakis}(2024)}]{Spaeh:2024:markov-chain-mixtures}%
  \BibitemOpen
  \bibfield  {author} {\bibinfo {author} {\bibfnamefont {F.}~\bibnamefont {Spaeh}}\ and\ \bibinfo {author} {\bibfnamefont {C.~E.}\ \bibnamefont {Tsourakakis}},\ }\bibfield  {title} {\bibinfo {title} {Markovletics: Methods and a novel application for learning continuous-time {M}arkov chain mixtures},\ }in\ \href {https://doi.org/10.1145/3589334.3645491} {\emph {\bibinfo {booktitle} {Proc. ACM Web Conf. 2024}}}\ (\bibinfo {year} {2024})\ pp.\ \bibinfo {pages} {4160--4171}\BibitemShut {NoStop}%
\bibitem [{\citenamefont {Hopfield}(1974)}]{Hopfield:1974:KineticProofreading}%
  \BibitemOpen
  \bibfield  {author} {\bibinfo {author} {\bibfnamefont {J.~J.}\ \bibnamefont {Hopfield}},\ }\bibfield  {title} {\bibinfo {title} {Kinetic proofreading: A new mechanism for reducing errors in biosynthetic processes requiring high specificity},\ }\href {https://doi.org/10.1073/pnas.71.10.4135} {\bibfield  {journal} {\bibinfo  {journal} {Proc. Natl. Acad. Sci. USA}\ }\textbf {\bibinfo {volume} {71}},\ \bibinfo {pages} {4135} (\bibinfo {year} {1974})}\BibitemShut {NoStop}%
\bibitem [{\citenamefont {Ninio}(1975)}]{Ninio:1975:KineticAmplification}%
  \BibitemOpen
  \bibfield  {author} {\bibinfo {author} {\bibfnamefont {J.}~\bibnamefont {Ninio}},\ }\bibfield  {title} {\bibinfo {title} {Kinetic amplification of enzyme discrimination},\ }\href {https://doi.org/10.1016/S0300-9084(75)80139-8} {\bibfield  {journal} {\bibinfo  {journal} {Biochimie}\ }\textbf {\bibinfo {volume} {57}},\ \bibinfo {pages} {587} (\bibinfo {year} {1975})}\BibitemShut {NoStop}%
\bibitem [{\citenamefont {McKeithan}(1995)}]{McKeithan:1995:KineticProofreading}%
  \BibitemOpen
  \bibfield  {author} {\bibinfo {author} {\bibfnamefont {T.~W.}\ \bibnamefont {McKeithan}},\ }\bibfield  {title} {\bibinfo {title} {Kinetic proofreading in {T}-cell receptor signal transduction},\ }\href {https://doi.org/10.1073/pnas.92.11.5042} {\bibfield  {journal} {\bibinfo  {journal} {Proc. Natl. Acad. Sci. U.S.A.}\ }\textbf {\bibinfo {volume} {92}},\ \bibinfo {pages} {5042} (\bibinfo {year} {1995})}\BibitemShut {NoStop}%
\bibitem [{\citenamefont {Murugan}\ \emph {et~al.}(2014)\citenamefont {Murugan}, \citenamefont {Huse},\ and\ \citenamefont {Leibler}}]{Murugan:2014:DiscriminatoryProofreading}%
  \BibitemOpen
  \bibfield  {author} {\bibinfo {author} {\bibfnamefont {A.}~\bibnamefont {Murugan}}, \bibinfo {author} {\bibfnamefont {D.~A.}\ \bibnamefont {Huse}},\ and\ \bibinfo {author} {\bibfnamefont {S.}~\bibnamefont {Leibler}},\ }\bibfield  {title} {\bibinfo {title} {Discriminatory proofreading regimes in nonequilibrium systems},\ }\href {https://doi.org/10.1103/PhysRevX.4.021016} {\bibfield  {journal} {\bibinfo  {journal} {Phys. Rev. X}\ }\textbf {\bibinfo {volume} {4}},\ \bibinfo {pages} {021016} (\bibinfo {year} {2014})}\BibitemShut {NoStop}%
\bibitem [{\citenamefont {Hartich}\ \emph {et~al.}(2015)\citenamefont {Hartich}, \citenamefont {Barato},\ and\ \citenamefont {Seifert}}]{Hartich:2015:NonequilibriumSensing}%
  \BibitemOpen
  \bibfield  {author} {\bibinfo {author} {\bibfnamefont {D.}~\bibnamefont {Hartich}}, \bibinfo {author} {\bibfnamefont {A.~C.}\ \bibnamefont {Barato}},\ and\ \bibinfo {author} {\bibfnamefont {U.}~\bibnamefont {Seifert}},\ }\bibfield  {title} {\bibinfo {title} {Nonequilibrium sensing and its analogy to kinetic proofreading},\ }\href {https://doi.org/10.1088/1367-2630/17/5/055026} {\bibfield  {journal} {\bibinfo  {journal} {New J. Phys.}\ }\textbf {\bibinfo {volume} {17}},\ \bibinfo {pages} {055026} (\bibinfo {year} {2015})}\BibitemShut {NoStop}%
\bibitem [{\citenamefont {Rao}\ and\ \citenamefont {Peliti}(2015)}]{Rao:2015:ThermodynamicsAccuracyProofreading}%
  \BibitemOpen
  \bibfield  {author} {\bibinfo {author} {\bibfnamefont {R.}~\bibnamefont {Rao}}\ and\ \bibinfo {author} {\bibfnamefont {L.}~\bibnamefont {Peliti}},\ }\bibfield  {title} {\bibinfo {title} {Thermodynamics of accuracy in kinetic proofreading: dissipation and efficiency trade-offs},\ }\href {https://doi.org/10.1088/1742-5468/2015/06/P06001} {\bibfield  {journal} {\bibinfo  {journal} {J. Stat. Mech.}\ }\textbf {\bibinfo {volume} {2015}},\ \bibinfo {pages} {P06001} (\bibinfo {year} {2015})}\BibitemShut {NoStop}%
\bibitem [{\citenamefont {Wong}\ \emph {et~al.}(2018)\citenamefont {Wong}, \citenamefont {Amir},\ and\ \citenamefont {Gunawardena}}]{Wong:2018:EnergySpeedAccuracy}%
  \BibitemOpen
  \bibfield  {author} {\bibinfo {author} {\bibfnamefont {F.}~\bibnamefont {Wong}}, \bibinfo {author} {\bibfnamefont {A.}~\bibnamefont {Amir}},\ and\ \bibinfo {author} {\bibfnamefont {J.}~\bibnamefont {Gunawardena}},\ }\bibfield  {title} {\bibinfo {title} {Energy-speed-accuracy relation in complex networks for biological discrimination},\ }\href {https://doi.org/10.1103/PhysRevE.98.012420} {\bibfield  {journal} {\bibinfo  {journal} {Phys. Rev. E}\ }\textbf {\bibinfo {volume} {98}},\ \bibinfo {pages} {012420} (\bibinfo {year} {2018})}\BibitemShut {NoStop}%
\bibitem [{\citenamefont {Ganti}\ \emph {et~al.}(2020)\citenamefont {Ganti}, \citenamefont {Lo}, \citenamefont {McAffee}, \citenamefont {Groves}, \citenamefont {Weiss},\ and\ \citenamefont {Chakraborty}}]{Ganti:2020:TCellSignaling}%
  \BibitemOpen
  \bibfield  {author} {\bibinfo {author} {\bibfnamefont {R.~S.}\ \bibnamefont {Ganti}}, \bibinfo {author} {\bibfnamefont {W.}~\bibnamefont {Lo}}, \bibinfo {author} {\bibfnamefont {D.~B.}\ \bibnamefont {McAffee}}, \bibinfo {author} {\bibfnamefont {J.~T.}\ \bibnamefont {Groves}}, \bibinfo {author} {\bibfnamefont {A.}~\bibnamefont {Weiss}},\ and\ \bibinfo {author} {\bibfnamefont {A.~K.}\ \bibnamefont {Chakraborty}},\ }\bibfield  {title} {\bibinfo {title} {How the t cell signaling network processes information to discriminate between self and agonist ligands},\ }\href {https://doi.org/10.1073/pnas.2008303117} {\bibfield  {journal} {\bibinfo  {journal} {Proc. Natl. Acad. Sci. U.S.A.}\ }\textbf {\bibinfo {volume} {117}},\ \bibinfo {pages} {26020} (\bibinfo {year} {2020})}\BibitemShut {NoStop}%
\bibitem [{\citenamefont {Berx}\ and\ \citenamefont {Proesmans}(2024)}]{Berx:2024:tradeoffs-thermodynamics-energy-relay}%
  \BibitemOpen
  \bibfield  {author} {\bibinfo {author} {\bibfnamefont {J.}~\bibnamefont {Berx}}\ and\ \bibinfo {author} {\bibfnamefont {K.}~\bibnamefont {Proesmans}},\ }\bibfield  {title} {\bibinfo {title} {Trade-offs and thermodynamics of energy-relay proofreading},\ }\href {https://doi.org/10.1098/rsif.2024.0232} {\bibfield  {journal} {\bibinfo  {journal} {J. R. Soc. Interface}\ }\textbf {\bibinfo {volume} {21}},\ \bibinfo {pages} {20240232} (\bibinfo {year} {2024})}\BibitemShut {NoStop}%
\bibitem [{\citenamefont {Berg}\ and\ \citenamefont {Purcell}(1977)}]{Berg:1977:PhysicsOfChemoreception}%
  \BibitemOpen
  \bibfield  {author} {\bibinfo {author} {\bibfnamefont {H.~C.}\ \bibnamefont {Berg}}\ and\ \bibinfo {author} {\bibfnamefont {E.~M.}\ \bibnamefont {Purcell}},\ }\bibfield  {title} {\bibinfo {title} {Physics of chemoreception},\ }\href {https://doi.org/10.1016/S0006-3495(77)85544-6} {\bibfield  {journal} {\bibinfo  {journal} {Biophys. J.}\ }\textbf {\bibinfo {volume} {20}},\ \bibinfo {pages} {193} (\bibinfo {year} {1977})}\BibitemShut {NoStop}%
\bibitem [{\citenamefont {Endres}\ and\ \citenamefont {Wingreen}(2009)}]{Endres:2009:MLE}%
  \BibitemOpen
  \bibfield  {author} {\bibinfo {author} {\bibfnamefont {R.~G.}\ \bibnamefont {Endres}}\ and\ \bibinfo {author} {\bibfnamefont {N.~S.}\ \bibnamefont {Wingreen}},\ }\bibfield  {title} {\bibinfo {title} {Maximum likelihood and the single receptor},\ }\href {https://doi.org/10.1103/PhysRevLett.103.158101} {\bibfield  {journal} {\bibinfo  {journal} {Phys. Rev. Lett.}\ }\textbf {\bibinfo {volume} {103}},\ \bibinfo {pages} {158101} (\bibinfo {year} {2009})}\BibitemShut {NoStop}%
\bibitem [{\citenamefont {Mora}\ and\ \citenamefont {Wingreen}(2010)}]{Mora:2010:MLE}%
  \BibitemOpen
  \bibfield  {author} {\bibinfo {author} {\bibfnamefont {T.}~\bibnamefont {Mora}}\ and\ \bibinfo {author} {\bibfnamefont {N.~S.}\ \bibnamefont {Wingreen}},\ }\bibfield  {title} {\bibinfo {title} {Limits of sensing temporal concentration changes by single cells},\ }\href {https://doi.org/10.1103/PhysRevLett.104.248101} {\bibfield  {journal} {\bibinfo  {journal} {Phys. Rev. Lett.}\ }\textbf {\bibinfo {volume} {104}},\ \bibinfo {pages} {248101} (\bibinfo {year} {2010})}\BibitemShut {NoStop}%
\bibitem [{\citenamefont {Mehta}\ and\ \citenamefont {Schwab}(2012)}]{Mehta:2012:Energetic}%
  \BibitemOpen
  \bibfield  {author} {\bibinfo {author} {\bibfnamefont {P.}~\bibnamefont {Mehta}}\ and\ \bibinfo {author} {\bibfnamefont {D.~J.}\ \bibnamefont {Schwab}},\ }\bibfield  {title} {\bibinfo {title} {Energetic costs of cellular computation},\ }\href {https://doi.org/10.1073/pnas.1207814109} {\bibfield  {journal} {\bibinfo  {journal} {Proc. Natl. Acad. Sci. U.S.A.}\ }\textbf {\bibinfo {volume} {109}},\ \bibinfo {pages} {17978} (\bibinfo {year} {2012})}\BibitemShut {NoStop}%
\bibitem [{\citenamefont {Govern}\ and\ \citenamefont {ten Wolde}(2014)}]{Govern:2014:EnergyDissipationNoise}%
  \BibitemOpen
  \bibfield  {author} {\bibinfo {author} {\bibfnamefont {C.~C.}\ \bibnamefont {Govern}}\ and\ \bibinfo {author} {\bibfnamefont {P.~R.}\ \bibnamefont {ten Wolde}},\ }\bibfield  {title} {\bibinfo {title} {Energy dissipation and noise correlations in biochemical sensing},\ }\href {https://doi.org/10.1103/PhysRevLett.113.258102} {\bibfield  {journal} {\bibinfo  {journal} {Phys. Rev. Lett.}\ }\textbf {\bibinfo {volume} {113}},\ \bibinfo {pages} {258102} (\bibinfo {year} {2014})}\BibitemShut {NoStop}%
\bibitem [{\citenamefont {Harvey}\ \emph {et~al.}(2023)\citenamefont {Harvey}, \citenamefont {Lahiri},\ and\ \citenamefont {Ganguli}}]{Harvey:2023:energy-accuracy-tradeoffs}%
  \BibitemOpen
  \bibfield  {author} {\bibinfo {author} {\bibfnamefont {S.~E.}\ \bibnamefont {Harvey}}, \bibinfo {author} {\bibfnamefont {S.}~\bibnamefont {Lahiri}},\ and\ \bibinfo {author} {\bibfnamefont {S.}~\bibnamefont {Ganguli}},\ }\bibfield  {title} {\bibinfo {title} {Universal energy-accuracy tradeoffs in nonequilibrium cellular sensing},\ }\href {https://doi.org/10.1103/PhysRevE.108.014403} {\bibfield  {journal} {\bibinfo  {journal} {Phys. Rev. E}\ }\textbf {\bibinfo {volume} {108}},\ \bibinfo {pages} {014403} (\bibinfo {year} {2023})}\BibitemShut {NoStop}%
\bibitem [{\citenamefont {Peres}\ and\ \citenamefont {Wootters}(1991)}]{Peres:1991:OptimalDetection}%
  \BibitemOpen
  \bibfield  {author} {\bibinfo {author} {\bibfnamefont {A.}~\bibnamefont {Peres}}\ and\ \bibinfo {author} {\bibfnamefont {W.~K.}\ \bibnamefont {Wootters}},\ }\bibfield  {title} {\bibinfo {title} {Optimal detection of quantum information},\ }\href {https://doi.org/10.1103/PhysRevLett.66.1119} {\bibfield  {journal} {\bibinfo  {journal} {Phys. Rev. Lett.}\ }\textbf {\bibinfo {volume} {66}},\ \bibinfo {pages} {1119} (\bibinfo {year} {1991})}\BibitemShut {NoStop}%
\bibitem [{\citenamefont {Massar}\ and\ \citenamefont {Popescu}(1995)}]{Massar:1995:OptimalExtraction}%
  \BibitemOpen
  \bibfield  {author} {\bibinfo {author} {\bibfnamefont {S.}~\bibnamefont {Massar}}\ and\ \bibinfo {author} {\bibfnamefont {S.}~\bibnamefont {Popescu}},\ }\bibfield  {title} {\bibinfo {title} {Optimal extraction of information from finite quantum ensembles},\ }\href {https://doi.org/10.1103/PhysRevLett.74.1259} {\bibfield  {journal} {\bibinfo  {journal} {Phys. Rev. Lett.}\ }\textbf {\bibinfo {volume} {74}},\ \bibinfo {pages} {1259} (\bibinfo {year} {1995})}\BibitemShut {NoStop}%
\bibitem [{\citenamefont {Gill}\ and\ \citenamefont {Massar}(2000)}]{Gill:2000:StateEstimation}%
  \BibitemOpen
  \bibfield  {author} {\bibinfo {author} {\bibfnamefont {R.~D.}\ \bibnamefont {Gill}}\ and\ \bibinfo {author} {\bibfnamefont {S.}~\bibnamefont {Massar}},\ }\bibfield  {title} {\bibinfo {title} {State estimation for large ensembles},\ }\href {https://doi.org/10.1103/PhysRevA.61.042312} {\bibfield  {journal} {\bibinfo  {journal} {Phys. Rev. A}\ }\textbf {\bibinfo {volume} {61}},\ \bibinfo {pages} {042312} (\bibinfo {year} {2000})}\BibitemShut {NoStop}%
\bibitem [{\citenamefont {Ekert}\ \emph {et~al.}(2002)\citenamefont {Ekert}, \citenamefont {Alves}, \citenamefont {Oi}, \citenamefont {Horodecki}, \citenamefont {Horodecki},\ and\ \citenamefont {Kwek}}]{Ekert:2002:DirectEstimations}%
  \BibitemOpen
  \bibfield  {author} {\bibinfo {author} {\bibfnamefont {A.~K.}\ \bibnamefont {Ekert}}, \bibinfo {author} {\bibfnamefont {C.~M.}\ \bibnamefont {Alves}}, \bibinfo {author} {\bibfnamefont {D.~K.~L.}\ \bibnamefont {Oi}}, \bibinfo {author} {\bibfnamefont {M.}~\bibnamefont {Horodecki}}, \bibinfo {author} {\bibfnamefont {P.}~\bibnamefont {Horodecki}},\ and\ \bibinfo {author} {\bibfnamefont {L.~C.}\ \bibnamefont {Kwek}},\ }\bibfield  {title} {\bibinfo {title} {Direct estimations of linear and nonlinear functionals of a quantum state},\ }\href {https://doi.org/10.1103/PhysRevLett.88.217901} {\bibfield  {journal} {\bibinfo  {journal} {Phys. Rev. Lett.}\ }\textbf {\bibinfo {volume} {88}},\ \bibinfo {pages} {217901} (\bibinfo {year} {2002})}\BibitemShut {NoStop}%
\bibitem [{\citenamefont {Ac{\'\i}n}\ \emph {et~al.}(2005)\citenamefont {Ac{\'\i}n}, \citenamefont {Bagan}, \citenamefont {Baig}, \citenamefont {Masanes},\ and\ \citenamefont {Mu{\~n}oz-Tapia}}]{Acin:2005:MultipleCopyDiscrimination}%
  \BibitemOpen
  \bibfield  {author} {\bibinfo {author} {\bibfnamefont {A.}~\bibnamefont {Ac{\'\i}n}}, \bibinfo {author} {\bibfnamefont {E.}~\bibnamefont {Bagan}}, \bibinfo {author} {\bibfnamefont {M.}~\bibnamefont {Baig}}, \bibinfo {author} {\bibfnamefont {L.}~\bibnamefont {Masanes}},\ and\ \bibinfo {author} {\bibfnamefont {R.}~\bibnamefont {Mu{\~n}oz-Tapia}},\ }\bibfield  {title} {\bibinfo {title} {Multiple-copy two-state discrimination with individual measurements},\ }\href {https://doi.org/10.1103/PhysRevA.71.032338} {\bibfield  {journal} {\bibinfo  {journal} {Phys. Rev. A}\ }\textbf {\bibinfo {volume} {71}},\ \bibinfo {pages} {032338} (\bibinfo {year} {2005})}\BibitemShut {NoStop}%
\bibitem [{\citenamefont {Seifert}(2005)}]{Seifert:2005:FT}%
  \BibitemOpen
  \bibfield  {author} {\bibinfo {author} {\bibfnamefont {U.}~\bibnamefont {Seifert}},\ }\bibfield  {title} {\bibinfo {title} {Entropy production along a stochastic trajectory and an integral fluctuation theorem},\ }\href {https://doi.org/10.1103/PhysRevLett.95.040602} {\bibfield  {journal} {\bibinfo  {journal} {Phys. Rev. Lett.}\ }\textbf {\bibinfo {volume} {95}},\ \bibinfo {pages} {040602} (\bibinfo {year} {2005})}\BibitemShut {NoStop}%
\bibitem [{\citenamefont {Barato}\ and\ \citenamefont {Seifert}(2015)}]{Barato:2015:UncRel}%
  \BibitemOpen
  \bibfield  {author} {\bibinfo {author} {\bibfnamefont {A.~C.}\ \bibnamefont {Barato}}\ and\ \bibinfo {author} {\bibfnamefont {U.}~\bibnamefont {Seifert}},\ }\bibfield  {title} {\bibinfo {title} {Thermodynamic uncertainty relation for biomolecular processes},\ }\href {https://doi.org/10.1103/PhysRevLett.114.158101} {\bibfield  {journal} {\bibinfo  {journal} {Phys. Rev. Lett.}\ }\textbf {\bibinfo {volume} {114}},\ \bibinfo {pages} {158101} (\bibinfo {year} {2015})}\BibitemShut {NoStop}%
\bibitem [{\citenamefont {Gingrich}\ \emph {et~al.}(2016)\citenamefont {Gingrich}, \citenamefont {Horowitz}, \citenamefont {Perunov},\ and\ \citenamefont {England}}]{Gingrich:2016:TUP}%
  \BibitemOpen
  \bibfield  {author} {\bibinfo {author} {\bibfnamefont {T.~R.}\ \bibnamefont {Gingrich}}, \bibinfo {author} {\bibfnamefont {J.~M.}\ \bibnamefont {Horowitz}}, \bibinfo {author} {\bibfnamefont {N.}~\bibnamefont {Perunov}},\ and\ \bibinfo {author} {\bibfnamefont {J.~L.}\ \bibnamefont {England}},\ }\bibfield  {title} {\bibinfo {title} {Dissipation bounds all steady-state current fluctuations},\ }\href {https://doi.org/10.1103/PhysRevLett.116.120601} {\bibfield  {journal} {\bibinfo  {journal} {Phys. Rev. Lett.}\ }\textbf {\bibinfo {volume} {116}},\ \bibinfo {pages} {120601} (\bibinfo {year} {2016})}\BibitemShut {NoStop}%
\bibitem [{\citenamefont {Garrahan}(2017)}]{Garrahan:2017:TUR}%
  \BibitemOpen
  \bibfield  {author} {\bibinfo {author} {\bibfnamefont {J.~P.}\ \bibnamefont {Garrahan}},\ }\bibfield  {title} {\bibinfo {title} {Simple bounds on fluctuations and uncertainty relations for first-passage times of counting observables},\ }\href {https://doi.org/10.1103/PhysRevE.95.032134} {\bibfield  {journal} {\bibinfo  {journal} {Phys. Rev. E}\ }\textbf {\bibinfo {volume} {95}},\ \bibinfo {pages} {032134} (\bibinfo {year} {2017})}\BibitemShut {NoStop}%
\bibitem [{\citenamefont {Shiraishi}\ \emph {et~al.}(2018)\citenamefont {Shiraishi}, \citenamefont {Funo},\ and\ \citenamefont {Saito}}]{Shiraishi:2018:SpeedLimit}%
  \BibitemOpen
  \bibfield  {author} {\bibinfo {author} {\bibfnamefont {N.}~\bibnamefont {Shiraishi}}, \bibinfo {author} {\bibfnamefont {K.}~\bibnamefont {Funo}},\ and\ \bibinfo {author} {\bibfnamefont {K.}~\bibnamefont {Saito}},\ }\bibfield  {title} {\bibinfo {title} {Speed limit for classical stochastic processes},\ }\href {https://link.aps.org/doi/10.1103/PhysRevLett.121.070601} {\bibfield  {journal} {\bibinfo  {journal} {Phys. Rev. Lett.}\ }\textbf {\bibinfo {volume} {121}},\ \bibinfo {pages} {070601} (\bibinfo {year} {2018})}\BibitemShut {NoStop}%
\bibitem [{\citenamefont {{Di Terlizzi}}\ and\ \citenamefont {Baiesi}(2019)}]{Terlizzi:2019:KUR}%
  \BibitemOpen
  \bibfield  {author} {\bibinfo {author} {\bibfnamefont {I.}~\bibnamefont {{Di Terlizzi}}}\ and\ \bibinfo {author} {\bibfnamefont {M.}~\bibnamefont {Baiesi}},\ }\bibfield  {title} {\bibinfo {title} {Kinetic uncertainty relation},\ }\href {https://doi.org/10.1088/1751-8121/aaee34} {\bibfield  {journal} {\bibinfo  {journal} {J. Phys. A: Math. Theor.}\ }\textbf {\bibinfo {volume} {52}},\ \bibinfo {pages} {02LT03} (\bibinfo {year} {2019})}\BibitemShut {NoStop}%
\bibitem [{\citenamefont {Hasegawa}(2020)}]{Hasegawa:2020:QTURPRL}%
  \BibitemOpen
  \bibfield  {author} {\bibinfo {author} {\bibfnamefont {Y.}~\bibnamefont {Hasegawa}},\ }\bibfield  {title} {\bibinfo {title} {Quantum thermodynamic uncertainty relation for continuous measurement},\ }\href {https://doi.org/10.1103/PhysRevLett.125.050601} {\bibfield  {journal} {\bibinfo  {journal} {Phys. Rev. Lett.}\ }\textbf {\bibinfo {volume} {125}},\ \bibinfo {pages} {050601} (\bibinfo {year} {2020})}\BibitemShut {NoStop}%
\bibitem [{\citenamefont {Hasegawa}(2023)}]{Hasegawa:2023:BulkBoundaryBoundNC}%
  \BibitemOpen
  \bibfield  {author} {\bibinfo {author} {\bibfnamefont {Y.}~\bibnamefont {Hasegawa}},\ }\bibfield  {title} {\bibinfo {title} {Unifying speed limit, thermodynamic uncertainty relation and {Heisenberg} principle via bulk-boundary correspondence},\ }\href {https://doi.org/10.1038/s41467-023-38074-8} {\bibfield  {journal} {\bibinfo  {journal} {Nat. Commun.}\ }\textbf {\bibinfo {volume} {14}},\ \bibinfo {pages} {2828} (\bibinfo {year} {2023})}\BibitemShut {NoStop}%
\bibitem [{\citenamefont {Devroye}\ \emph {et~al.}(1996)\citenamefont {Devroye}, \citenamefont {Gy{\"o}rfi},\ and\ \citenamefont {Lugosi}}]{Devroye:1996:pattern-recognition}%
  \BibitemOpen
  \bibfield  {author} {\bibinfo {author} {\bibfnamefont {L.}~\bibnamefont {Devroye}}, \bibinfo {author} {\bibfnamefont {L.}~\bibnamefont {Gy{\"o}rfi}},\ and\ \bibinfo {author} {\bibfnamefont {G.}~\bibnamefont {Lugosi}},\ }\href {https://doi.org/10.1007/978-1-4612-0711-5} {\emph {\bibinfo {title} {A Probabilistic Theory of Pattern Recognition}}},\ \bibinfo {edition} {1st}\ ed.,\ \bibinfo {series} {Stochastic Modelling and Applied Probability}, Vol.~\bibinfo {volume} {31}\ (\bibinfo {address} {New York, NY},\ \bibinfo {year} {1996})\BibitemShut {NoStop}%
\bibitem [{\citenamefont {Nielsen}(2014)}]{Nielsen:2014:Generalized}%
  \BibitemOpen
  \bibfield  {author} {\bibinfo {author} {\bibfnamefont {F.}~\bibnamefont {Nielsen}},\ }\bibfield  {title} {\bibinfo {title} {Generalized {Bhattacharyya} and {Chernoff} upper bounds on {B}ayes error using quasi-arithmetic means},\ }\href {https://doi.org/10.1016/j.patrec.2014.01.002} {\bibfield  {journal} {\bibinfo  {journal} {Pattern Recognit. Lett.}\ }\textbf {\bibinfo {volume} {42}},\ \bibinfo {pages} {25} (\bibinfo {year} {2014})}\BibitemShut {NoStop}%
\bibitem [{Sup()}]{Supp:2026:ThermoClassifier}%
  \BibitemOpen
  \href@noop {} {}\bibinfo {note} {See Supplemental Material for details of calculations, which includes \cite{Sason:2016:DivIneqReview,Pedregosa:2011:scikit-learn}.}\BibitemShut {Stop}%
\bibitem [{\citenamefont {Pagare}\ and\ \citenamefont {Lu}(2024)}]{Pagare:2024:MpembaLikeSensor}%
  \BibitemOpen
  \bibfield  {author} {\bibinfo {author} {\bibfnamefont {A.}~\bibnamefont {Pagare}}\ and\ \bibinfo {author} {\bibfnamefont {Z.}~\bibnamefont {Lu}},\ }\bibfield  {title} {\bibinfo {title} {Mpemba-like sensory withdrawal effect},\ }\href {https://doi.org/10.1103/PRXLife.2.043019} {\bibfield  {journal} {\bibinfo  {journal} {PRX Life}\ }\textbf {\bibinfo {volume} {2}},\ \bibinfo {pages} {043019} (\bibinfo {year} {2024})}\BibitemShut {NoStop}%
\bibitem [{\citenamefont {Sason}\ and\ \citenamefont {Verd\'u}(2016)}]{Sason:2016:DivIneqReview}%
  \BibitemOpen
  \bibfield  {author} {\bibinfo {author} {\bibfnamefont {I.}~\bibnamefont {Sason}}\ and\ \bibinfo {author} {\bibfnamefont {S.}~\bibnamefont {Verd\'u}},\ }\bibfield  {title} {\bibinfo {title} {$f$-divergence inequalities},\ }\href {https://doi.org/10.1109/TIT.2016.2603151} {\bibfield  {journal} {\bibinfo  {journal} {IEEE Trans. Inf. Theory}\ }\textbf {\bibinfo {volume} {62}},\ \bibinfo {pages} {5973} (\bibinfo {year} {2016})}\BibitemShut {NoStop}%
\bibitem [{\citenamefont {Pedregosa}\ \emph {et~al.}(2011)\citenamefont {Pedregosa}, \citenamefont {Varoquaux}, \citenamefont {Gramfort}, \citenamefont {Michel}, \citenamefont {Thirion}, \citenamefont {Grisel}, \citenamefont {Blondel}, \citenamefont {Prettenhofer}, \citenamefont {Weiss}, \citenamefont {Dubourg}, \citenamefont {Vanderplas}, \citenamefont {Passos}, \citenamefont {Cournapeau}, \citenamefont {Brucher}, \citenamefont {Perrot},\ and\ \citenamefont {Duchesnay}}]{Pedregosa:2011:scikit-learn}%
  \BibitemOpen
  \bibfield  {author} {\bibinfo {author} {\bibfnamefont {F.}~\bibnamefont {Pedregosa}}, \bibinfo {author} {\bibfnamefont {G.}~\bibnamefont {Varoquaux}}, \bibinfo {author} {\bibfnamefont {A.}~\bibnamefont {Gramfort}}, \bibinfo {author} {\bibfnamefont {V.}~\bibnamefont {Michel}}, \bibinfo {author} {\bibfnamefont {B.}~\bibnamefont {Thirion}}, \bibinfo {author} {\bibfnamefont {O.}~\bibnamefont {Grisel}}, \bibinfo {author} {\bibfnamefont {M.}~\bibnamefont {Blondel}}, \bibinfo {author} {\bibfnamefont {P.}~\bibnamefont {Prettenhofer}}, \bibinfo {author} {\bibfnamefont {R.}~\bibnamefont {Weiss}}, \bibinfo {author} {\bibfnamefont {V.}~\bibnamefont {Dubourg}}, \bibinfo {author} {\bibfnamefont {J.}~\bibnamefont {Vanderplas}}, \bibinfo {author} {\bibfnamefont {A.}~\bibnamefont {Passos}}, \bibinfo {author} {\bibfnamefont {D.}~\bibnamefont {Cournapeau}}, \bibinfo {author} {\bibfnamefont {M.}~\bibnamefont {Brucher}}, \bibinfo {author} {\bibfnamefont {M.}~\bibnamefont {Perrot}},\ and\ \bibinfo {author} {\bibfnamefont
  {E.}~\bibnamefont {Duchesnay}},\ }\bibfield  {title} {\bibinfo {title} {Scikit-learn: Machine learning in {P}ython},\ }\href {https://www.jmlr.org/papers/v12/pedregosa11a.html} {\bibfield  {journal} {\bibinfo  {journal} {J. Mach. Learn. Res.}\ }\textbf {\bibinfo {volume} {12}},\ \bibinfo {pages} {2825} (\bibinfo {year} {2011})}\BibitemShut {NoStop}%
\bibitem [{\citenamefont {Lehmann}\ and\ \citenamefont {Casella}(1998)}]{Lehmann:1998:PointEstimation}%
  \BibitemOpen
  \bibfield  {author} {\bibinfo {author} {\bibfnamefont {E.~L.}\ \bibnamefont {Lehmann}}\ and\ \bibinfo {author} {\bibfnamefont {G.}~\bibnamefont {Casella}},\ }\href {https://doi.org/10.1007/b98854} {\emph {\bibinfo {title} {Theory of Point Estimation}}},\ \bibinfo {edition} {2nd}\ ed.,\ Springer Texts in Statistics\ (\bibinfo {address} {New York, NY},\ \bibinfo {year} {1998})\BibitemShut {NoStop}%
\bibitem [{\citenamefont {Le~Cam}(1973)}]{LeCam:1973:dimensionality}%
  \BibitemOpen
  \bibfield  {author} {\bibinfo {author} {\bibfnamefont {L.}~\bibnamefont {Le~Cam}},\ }\bibfield  {title} {\bibinfo {title} {Convergence of estimates under dimensionality restrictions},\ }\href {https://doi.org/10.1214/aos/1193342380} {\bibfield  {journal} {\bibinfo  {journal} {Ann. Statist.}\ }\textbf {\bibinfo {volume} {1}},\ \bibinfo {pages} {38} (\bibinfo {year} {1973})}\BibitemShut {NoStop}%
\bibitem [{\citenamefont {Raginsky}(2013)}]{Raginsky:2013:SLT}%
  \BibitemOpen
  \bibfield  {author} {\bibinfo {author} {\bibfnamefont {M.}~\bibnamefont {Raginsky}},\ }\href@noop {} {\bibinfo {title} {Statistical learning theory lecture notes}},\ \bibinfo {howpublished} {\url{https://maxim.ece.illinois.edu/teaching/fall13/schedule.html}} (\bibinfo {year} {2013}),\ \bibinfo {note} {university of Illinois at Urbana-Champaign}\BibitemShut {NoStop}%
\bibitem [{\citenamefont {Wainwright}(2019)}]{Wainwright:2019:HighDimensionalStatistics}%
  \BibitemOpen
  \bibfield  {author} {\bibinfo {author} {\bibfnamefont {M.~J.}\ \bibnamefont {Wainwright}},\ }\href {https://doi.org/10.1017/9781108627771} {\emph {\bibinfo {title} {High-Dimensional Statistics: A Non-Asymptotic Viewpoint}}},\ Cambridge Series in Statistical and Probabilistic Mathematics\ (\bibinfo {address} {Cambridge},\ \bibinfo {year} {2019})\BibitemShut {NoStop}%
\bibitem [{\citenamefont {Jarzynski}(2011)}]{Jarzynski:2011:Irreversibility}%
  \BibitemOpen
  \bibfield  {author} {\bibinfo {author} {\bibfnamefont {C.}~\bibnamefont {Jarzynski}},\ }\bibfield  {title} {\bibinfo {title} {Equalities and inequalities: Irreversibility and the second law of thermodynamics at the nanoscale},\ }\href {https://doi.org/10.1146/annurev-conmatphys-062910-140506} {\bibfield  {journal} {\bibinfo  {journal} {Annu. Rev. Condens. Matter Phys.}\ }\textbf {\bibinfo {volume} {2}},\ \bibinfo {pages} {329} (\bibinfo {year} {2011})}\BibitemShut {NoStop}%
\bibitem [{\citenamefont {Rold\'an}\ \emph {et~al.}(2015)\citenamefont {Rold\'an}, \citenamefont {Neri}, \citenamefont {D\"orpinghaus}, \citenamefont {Meyr},\ and\ \citenamefont {J\"ulicher}}]{Roldan:2015:ArrowTimeDecision}%
  \BibitemOpen
  \bibfield  {author} {\bibinfo {author} {\bibfnamefont {E.}~\bibnamefont {Rold\'an}}, \bibinfo {author} {\bibfnamefont {I.}~\bibnamefont {Neri}}, \bibinfo {author} {\bibfnamefont {M.}~\bibnamefont {D\"orpinghaus}}, \bibinfo {author} {\bibfnamefont {H.}~\bibnamefont {Meyr}},\ and\ \bibinfo {author} {\bibfnamefont {F.}~\bibnamefont {J\"ulicher}},\ }\bibfield  {title} {\bibinfo {title} {Decision making in the arrow of time},\ }\href {https://doi.org/10.1103/PhysRevLett.115.250602} {\bibfield  {journal} {\bibinfo  {journal} {Phys. Rev. Lett.}\ }\textbf {\bibinfo {volume} {115}},\ \bibinfo {pages} {250602} (\bibinfo {year} {2015})}\BibitemShut {NoStop}%
\bibitem [{\citenamefont {Seif}\ \emph {et~al.}(2021)\citenamefont {Seif}, \citenamefont {Hafezi},\ and\ \citenamefont {Jarzynski}}]{Seif:2021:ThermoArrowTime}%
  \BibitemOpen
  \bibfield  {author} {\bibinfo {author} {\bibfnamefont {A.}~\bibnamefont {Seif}}, \bibinfo {author} {\bibfnamefont {M.}~\bibnamefont {Hafezi}},\ and\ \bibinfo {author} {\bibfnamefont {C.}~\bibnamefont {Jarzynski}},\ }\bibfield  {title} {\bibinfo {title} {Machine learning the thermodynamic arrow of time},\ }\href {https://doi.org/10.1038/s41567-020-1018-2} {\bibfield  {journal} {\bibinfo  {journal} {Nat. Phys.}\ }\textbf {\bibinfo {volume} {17}},\ \bibinfo {pages} {105} (\bibinfo {year} {2021})}\BibitemShut {NoStop}%
\bibitem [{\citenamefont {Wootters}(1981)}]{Wootters:1981:StatDist}%
  \BibitemOpen
  \bibfield  {author} {\bibinfo {author} {\bibfnamefont {W.~K.}\ \bibnamefont {Wootters}},\ }\bibfield  {title} {\bibinfo {title} {Statistical distance and {Hilbert} space},\ }\href {https://doi.org/10.1103/PhysRevD.23.357} {\bibfield  {journal} {\bibinfo  {journal} {Phys. Rev. D}\ }\textbf {\bibinfo {volume} {23}},\ \bibinfo {pages} {357} (\bibinfo {year} {1981})}\BibitemShut {NoStop}%
\bibitem [{\citenamefont {Ito}(2018)}]{Ito:2018:InfoGeo}%
  \BibitemOpen
  \bibfield  {author} {\bibinfo {author} {\bibfnamefont {S.}~\bibnamefont {Ito}},\ }\bibfield  {title} {\bibinfo {title} {Stochastic thermodynamic interpretation of information geometry},\ }\href {https://doi.org/10.1103/PhysRevLett.121.030605} {\bibfield  {journal} {\bibinfo  {journal} {Phys. Rev. Lett.}\ }\textbf {\bibinfo {volume} {121}},\ \bibinfo {pages} {030605} (\bibinfo {year} {2018})}\BibitemShut {NoStop}%
\bibitem [{\citenamefont {Nicholson}\ \emph {et~al.}(2020)\citenamefont {Nicholson}, \citenamefont {Garcia-Pintos}, \citenamefont {del Campo},\ and\ \citenamefont {Green}}]{Nicholson:2020:TIUncRel}%
  \BibitemOpen
  \bibfield  {author} {\bibinfo {author} {\bibfnamefont {S.~B.}\ \bibnamefont {Nicholson}}, \bibinfo {author} {\bibfnamefont {L.~P.}\ \bibnamefont {Garcia-Pintos}}, \bibinfo {author} {\bibfnamefont {A.}~\bibnamefont {del Campo}},\ and\ \bibinfo {author} {\bibfnamefont {J.~R.}\ \bibnamefont {Green}},\ }\bibfield  {title} {\bibinfo {title} {Time-information uncertainty relations in thermodynamics},\ }\href {https://doi.org/10.1038/s41567-020-0981-y} {\bibfield  {journal} {\bibinfo  {journal} {Nat. Phys.}\ }\textbf {\bibinfo {volume} {16}},\ \bibinfo {pages} {1211} (\bibinfo {year} {2020})}\BibitemShut {NoStop}%
\bibitem [{\citenamefont {Nishiyama}\ and\ \citenamefont {Hasegawa}(2026)}]{Nishiyama:2026:TemporalFisherPRE}%
  \BibitemOpen
  \bibfield  {author} {\bibinfo {author} {\bibfnamefont {T.}~\bibnamefont {Nishiyama}}\ and\ \bibinfo {author} {\bibfnamefont {Y.}~\bibnamefont {Hasegawa}},\ }\bibfield  {title} {\bibinfo {title} {Unified speed limits in classical and quantum dynamics via temporal {F}isher information},\ }\href {https://doi.org/10.1103/x95d-fhpq} {\bibfield  {journal} {\bibinfo  {journal} {Phys. Rev. E}\ }\textbf {\bibinfo {volume} {114}},\ \bibinfo {pages} {014120} (\bibinfo {year} {2026})}\BibitemShut {NoStop}%
\bibitem [{\citenamefont {Hasegawa}\ and\ \citenamefont {Nishiyama}(2024)}]{Hasegawa:2024:ConcentrationIneqPRL}%
  \BibitemOpen
  \bibfield  {author} {\bibinfo {author} {\bibfnamefont {Y.}~\bibnamefont {Hasegawa}}\ and\ \bibinfo {author} {\bibfnamefont {T.}~\bibnamefont {Nishiyama}},\ }\bibfield  {title} {\bibinfo {title} {Thermodynamic concentration inequalities and trade-off relations},\ }\href {https://doi.org/10.1103/PhysRevLett.133.247101} {\bibfield  {journal} {\bibinfo  {journal} {Phys. Rev. Lett.}\ }\textbf {\bibinfo {volume} {133}},\ \bibinfo {pages} {247101} (\bibinfo {year} {2024})}\BibitemShut {NoStop}%
\end{thebibliography}
\end{document}


\title{Supplementary Material for ``Thermodynamics of Classifiers''}

\author{Yoshihiko Hasegawa}
\email{hasegawa@biom.t.u-tokyo.ac.jp}
\affiliation{Department of Electrical Engineering and Information Systems, Graduate School of Engineering, The University of Tokyo,
Tokyo 113-8656, Japan}

\maketitle
This supplementary material describes the calculations introduced in the main text. The numbers of the equations and the figures are prefixed with S (e.g., Eq.~(S1) or Fig.~S1). Numbers without this prefix (e.g., Eq.~(1) or Fig.~1) refer to items in the main text.

\section{Derivation of the lower bound on the Bayes error\label{sec:error_lowerbound_derivation}}

We show the detailed derivation of the result presented in Eq.~\resultI{}. 
From the Bayes classifier given in Eq.~\BayesUerrorUclassical{}, 
it is given by
\begin{align}
P^{\mathrm{Bayes}}_{\mathrm{err}}=\sum_{x}\min\left[P(x\mid0)\pi_{0},P(x\mid1)\pi_{1}\right].
    \label{eq:Bayes_error}
\end{align}
We follow Refs.~\cite{Devroye:1996:pattern-recognition,Nielsen:2014:Generalized} to derive a lower bound on the Bayes error [Eq.~\eqref{eq:Bayes_error}] in terms of the Bhattacharyya coefficient. 
Recall that the min function can be represented by
\begin{align}
    \min (a, b)=\frac{a+b}{2}-\frac{1}{2}|b-a|.
    \label{eq:min_as_abs}
\end{align}
Let $\mathrm{TV}(\mathfrak{p},\mathfrak{q})$ be the total variation distance between general distributions $\mathfrak{p}$ and $\mathfrak{q}$:
\begin{align}
    \mathrm{TV}(\mathfrak{p},\mathfrak{q})\equiv\frac{1}{2}\sum_{x}|\mathfrak{p}(x)-\mathfrak{q}(x)|.
    \label{eq:TV_dist_def}
\end{align}
Using Eq.~\eqref{eq:min_as_abs}, the Bayes error is represented by \cite{Nielsen:2014:Generalized}
\begin{align}
    P^{\mathrm{Bayes}}_{\mathrm{err}}&=\frac{1}{2}-\frac{1}{2}\sum_{x}\left|P(x\mid y=0)\pi_{0}-P(x\mid y=1)\pi_{1}\right|\nonumber\\&=\frac{1}{2}-\mathrm{TV}\left(P(x\mid y=0)\pi_{0},P(x\mid y=1)\pi_{1}\right).
    \label{eq:Perr_min_TV}
\end{align}
Next, we consider a lower bound on the Bayes error using the representation in Eq.~\eqref{eq:Perr_min_TV}. 
Let $\mathrm{Bhat}(\mathfrak{p},\mathfrak{q})$ be the Bhattacharyya coefficient:
\begin{align}
    \mathrm{Bhat}(\mathfrak{p},\mathfrak{q})\equiv\sum_{x}\sqrt{\mathfrak{p}(x)\mathfrak{q}(x)}.
    \label{eq:Bhat_def}
\end{align}
It is known that the total variation distance is bounded from above by the Bhattacharyya coefficient \cite{LeCam:1973:Convergence,Sason:2016:DivIneqReview}:
\begin{align}
    \mathrm{TV}(\mathfrak{p},\mathfrak{q})\leq\sqrt{1-\mathrm{Bhat}(\mathfrak{p},\mathfrak{q})^{2}}.
    \label{eq:TV_and_Bhat}
\end{align}
Note that $\sum_{x}P(x\mid y)\pi_{y}\le1$, so $P(x\mid y)\pi_y$ is not normalized, and an extra factor of $\pi_0\pi_1$ will appear in front of the Bhattacharyya coefficient.
Using Eq.~\eqref{eq:TV_and_Bhat} in Eq.~\eqref{eq:Perr_min_TV}, the Bayes error has the following lower bound \cite{Devroye:1996:pattern-recognition,Nielsen:2014:Generalized} (Chapter 3 of Ref.~\cite{Devroye:1996:pattern-recognition}):
\begin{align}
    P^{\mathrm{Bayes}}_{\mathrm{err}}\ge\frac{1}{2}-\frac{1}{2}\sqrt{1-4\pi_{0}\pi_{1}\mathrm{Bhat}\left(P(x\mid0),P(x\mid1)\right)^{2}}.
    \label{eq:Bayes_Batt_lowerbound}
\end{align}
We use an inequality for the $p$-norm of a vector to obtain a lower bound on Eq.~\eqref{eq:Bayes_Batt_lowerbound}. 
For an $n$-dimensional vector $u$, the $p$-norm is defined by $\|u\|_p=\left(\sum_{k=1}^n\left|u_k\right|^p\right)^{1 / p}$. For $0<p<q$, it is known that $\|u\|_p \geq\|u\|_q$. 
Using this relation in Eq.~\eqref{eq:Bayes_Batt_lowerbound}, the Bayes error is bounded from below by
\begin{align}
    P^{\mathrm{Bayes}}_{\mathrm{err}}\ge\frac{1}{2}-\frac{1}{2}\sqrt{1-4\pi_{0}\pi_{1}\sum_{x}P(x\mid0)P(x\mid1)}.
    \label{eq:Perr_min_lowerbound_classical}
\end{align}

Since the state at time $t=\tau$ is used for classification, $\sum_{x}P(x\mid0)P(x\mid1)$ in Eq.~\eqref{eq:Perr_min_lowerbound_classical} should be replaced by $\sum_{\mu}P^{(0)}(\mu;\tau)P^{(1)}(\mu;\tau)$
when considering the scenario of classification based on the final state $z(\tau)$. 
Here, $P^{(y)}(\mu;\tau)$ is the probability distribution of the Markov process at time $t=\tau$ for the class label $Y=y$. 
We wish to find a lower bound on $\sum_{\mu}P^{(0)}(\mu;\tau)P^{(1)}(\mu;\tau)$.
Let $O(\mu)$ be an observable of state $\mu$ of a Markov process. 
We define the expectation of $O(\mu)$ as follows:
\begin{align}
   \mathbb{E}_{P(\mu;t)}[O]\equiv\sum_{\mu}O(\mu)P(\mu;t).
   \label{eq:O_expectation}
\end{align}
Then, according to the Cram\'er-Rao inequality, the following relation holds:
\begin{align}
    \frac{\mathrm{Var}_{P(\nu;t)}[O]}{\left(\partial_{t}\mathbb{E}_{P(\nu;t)}[O]\right)^{2}}\ge\frac{1}{\mathcal{I}(t)},
    \label{eq:CrameRao}
\end{align}
where $\mathrm{Var}_{P(\mu;t)}[O]$ is the variance and $\mathcal{I}(t)$ is the temporal Fisher information defined by
\begin{align}
    \mathcal{I}(t)&\equiv\sum_{\mu}P(\mu;t)\left(\frac{d}{dt}\ln P(\mu;t)\right)^{2}=\sum_{\mu}\frac{(\partial_{t}P(\mu;t))^{2}}{P(\mu;t)}.
    \label{eq:Fisher_information_def}
\end{align}
The temporal Fisher information
quantifies the infinitesimal distance in probability space \cite{Wootters:1981:StatDist} and
plays an important role in thermodynamic trade-off relations \cite{Ito:2018:InfoGeo,Nicholson:2020:TIUncRel}.
It is known that the following relation holds \cite{Nishiyama:2026:TemporalFisherPRE}:
\begin{align}
    \mathcal{I}(t)\le\frac{\Sigma(t)}{2t^{2}},\,\,\,\mathcal{I}(t)\le\frac{\mathcal{A}(t)}{t^{2}},\label{eq:It_upperbound}
\end{align}
where $\Sigma(t)$ and $\mathcal{A}(t)$ are the time-integrated entropy production and time-integrated dynamical activity, defined in Eqs.~\EPUdef{} and \integratedUactivityUdef{}, respectively. 
Note that the entropy production bound in Eq.~\eqref{eq:It_upperbound} holds for Langevin dynamics, whereas the dynamical activity version does not.

Next, we consider the combined Markov process in the enlarged space, as explained above in the Methods section. 
We consider two different Markov processes here. 
Let us consider the observable $O^{\oplus}(\mu^{(0)},\mu^{(1)})$, which is defined in the combined Markov process.
Since the combined Markov process comprises two independent Markov processes, $O^{\oplus}$ takes two arguments, corresponding to the states in the two Markov processes. 
The expectation of $O^\oplus$ is given by
\begin{align}
   \mathbb{E}_{P^{\oplus}(\mu^{(0)},\mu^{(1)};\tau)}[O^{\oplus}]&=\sum_{\mu^{(0)},\mu^{(1)}}O^{\oplus}(\mu^{(0)},\mu^{(1)})P^{\oplus}(\mu^{(0)},\mu^{(1)};\tau)\nonumber\\&=\sum_{\mu^{(0)},\mu^{(1)}}O^{\oplus}(\mu^{(0)},\mu^{(1)})P^{(0)}(\mu^{(0)};\tau)P^{(1)}(\mu^{(1)};\tau).
   \label{eq:O_expectation2}
\end{align}
Specifically, we employ the following observable:
\begin{align}
    \Omega^{\oplus}(\mu^{(0)},\mu^{(1)})=\begin{cases}
1 & \mu^{(0)}=\mu^{(1)}\\
0 & \mathrm{otherwise}
\end{cases}.
\label{eq:Oplus_def}
\end{align}
The observable expressed by Eq.~\eqref{eq:Oplus_def} is a quantity that returns $1$ if the final states of the two Markov processes match and $0$ otherwise.
This is reminiscent of the swap operator considered in multi-copy measurements \cite{Ekert:2002:DirectEstimations}. 
We emphasize that the two Markov processes are not coupled; however, the observable defined in Eq.~\eqref{eq:Oplus_def} is coupled.
The expectation and variance of $\Omega^\oplus$ are given by
\begin{align}
    \mathbb{E}_{P^{\oplus}(\mu^{(0)},\mu^{(1)};\tau)}[\Omega^{\oplus}]&=\alpha(\tau),\label{eq:E_Omega}\\\mathrm{Var}_{P^{\oplus}(\mu^{(0)},\mu^{(1)};\tau)}[\Omega^{\oplus}]&=\alpha(\tau)-\alpha(\tau)^{2},\label{eq:Var_Omega}
\end{align}
where
\begin{align}
    \alpha(\tau)\equiv\sum_{\mu}P^{(0)}(\mu;\tau)P^{(1)}(\mu;\tau).
    \label{eq:alpha_tau_def}
\end{align} 
Substituting Eqs.~\eqref{eq:E_Omega} and \eqref{eq:Var_Omega} into Eq.~\eqref{eq:CrameRao}, we obtain
\begin{align}
    \frac{\alpha(\tau)-\alpha(\tau)^{2}}{\left(\partial_{\tau}\alpha(\tau)\right)^{2}}\ge\frac{1}{\mathcal{I}^\oplus(\tau)}.
    \label{eq:alpha_ode}
\end{align}
Note that we consider the Cram\'er-Rao inequality in the combined Markov process. 
Equation~\eqref{eq:alpha_ode} is expressed as
\begin{align}
    \left|\frac{d}{d\tau}\arcsin(2\alpha(\tau)-1)\right|\le\sqrt{\mathcal{I}^\oplus(\tau)}.
    \label{eq:alpha_ode2}
\end{align}
Solving Eq.~\eqref{eq:alpha_ode2} with the condition $\alpha(0)=1$, we obtain
\begin{align}
    \alpha(\tau) \geq \cos\left(\frac{1}{2} \int_0^\tau \sqrt{\mathcal{I}^\oplus(t)} dt\right)^2,
    \label{eq:alphatau_bound}
\end{align}
which holds within the range $0\le\frac{1}{2}\int^{\tau}_{0}\sqrt{\mathcal{I}^\oplus(t)}dt\le\frac{\pi}{2}$. 
From Eq.~\eqref{eq:Perr_min_lowerbound_classical}, the Bayes error is
\begin{align}
    P^{\mathrm{Bayes}}_{\mathrm{err}}&\ge\frac{1}{2}-\frac{1}{2}\sqrt{1-4\pi_{0}\pi_{1}\alpha(\tau)}.
    \label{eq:Bayes_error_alphatau}
\end{align}
Substituting Eq.~\eqref{eq:alphatau_bound} into Eq.~\eqref{eq:Bayes_error_alphatau}, we obtain
\begin{align}
    P_\mathrm{err}^\mathrm{ave} \ge P^{\mathrm{Bayes}}_{\mathrm{err}}&\ge\frac{1}{2}-\frac{1}{2}\sqrt{1-4\pi_{0}\pi_{1}\cos\left(\frac{1}{2}\int^{\tau}_{0}\sqrt{\mathcal{I}^\oplus(t)}dt\right)^{2}}.
    \label{eq:Perr_min_It_bound}
\end{align}
Since $(\cos x)^2$ is a decreasing function for $0\le x \le \pi /2$, using Eq.~\eqref{eq:It_upperbound}, we obtain
\begin{align}
    P_\mathrm{err}^\mathrm{ave} \ge P^{\mathrm{Bayes}}_{\mathrm{err}}\ge&\frac{1}{2}-\frac{1}{2}\sqrt{1-4\pi_{0}\pi_{1}\cos\left(\frac{1}{2\sqrt{2}}\int^{\tau}_{0}\frac{\sqrt{\Sigma^{\oplus}(t)}}{t}dt\right)^{2}},\label{eq:Bayes_error_bound_Sigma_p0p1}\\P_\mathrm{err}^\mathrm{ave} \ge P^{\mathrm{Bayes}}_{\mathrm{err}}\ge&\frac{1}{2}-\frac{1}{2}\sqrt{1-4\pi_{0}\pi_{1}\cos\left(\frac{1}{2}\int^{\tau}_{0}\frac{\sqrt{\mathcal{A}^{\oplus}(t)}}{t}dt\right)^{2}},
    \label{eq:Bayes_error_bound_A_p0p1}
\end{align}
where
$\Sigma^{\oplus}(t) = \Sigma^{(0)}(t) + \Sigma^{(1)}(t)$ and 
$\mathcal{A}^{\oplus}(t)=\mathcal{A}^{(0)}(t)+\mathcal{A}^{(1)}(t)$.
Equations~\eqref{eq:Bayes_error_bound_Sigma_p0p1} and \eqref{eq:Bayes_error_bound_A_p0p1} hold within the ranges given in
\begin{align}
0 \leq \frac{1}{2 \sqrt{2}} \int_0^\tau \frac{\sqrt{\Sigma^{\oplus}(t)}}{t} d t \leq \frac{\pi}{2},\label{eq:Sigma_oplus_range}\\
0 \leq \frac{1}{2} \int_0^\tau \frac{\sqrt{\mathcal{A}^{\oplus}(t)}}{t} d t \leq \frac{\pi}{2}.
\label{eq:mathcalA_oplus_range}
\end{align}
Equations~\eqref{eq:Bayes_error_bound_Sigma_p0p1} and \eqref{eq:Bayes_error_bound_A_p0p1} are lower bounds on the Bayes error for general prior distributions $\pi_0$ and $\pi_1$. 
When we choose the uniform prior $\pi_0=\pi_1=1/2$, we obtain Eqs.~\resultI{} and \resultII{} in the main text. 

The inequality in Eq.~\resultII{} also holds for trajectory-based classification. 
In the combined Markov process, 
the relation in Ref.~\cite{Terlizzi:2019:KUR} holds:
\begin{align}
    \frac{\mathrm{Var}\left[N^{\oplus}\right]}{\tau^{2}\left(\partial_{\tau}\mathbb{E}\left[N^{\oplus}\right]\right)^{2}}\geq\frac{1}{\mathcal{A}^{\oplus}(\tau)}.
    \label{eq:KUR}
\end{align}
Here, $N^\oplus(\Gamma)$ is an arbitrary trajectory observable defined in the combined Markov process, and the expectation and variance are evaluated with respect to the path probability. 
To derive Eq.~\resultII{} for trajectory-based classification, we employ the following observable for $N^\oplus$ in Eq.~\eqref{eq:KUR}:
\begin{align}
    \Upsilon^{\oplus}\left(\Gamma^{\oplus}\right)\equiv\begin{cases}
1 & N_\circ\left(\Gamma^{(0)}\right)=N_\circ\left(\Gamma^{(1)}\right)\\
0 & \text{otherwise}
\end{cases}.
\label{eq:Upsilon_def}
\end{align}
Since Eq.~\eqref{eq:Upsilon_def} is identical to $\Omega^\oplus$ in Eq.~\eqref{eq:Oplus_def}, the subsequent calculations are the same. 
Let us define
\begin{align}
    \beta(\tau)\equiv\sum_{n}P(N_\circ(\Gamma^{(0)})=n)P(N_\circ(\Gamma^{(1)})=n).
    \label{eq:zetatau_def}
\end{align}
Then, Eq.~\eqref{eq:KUR} yields
\begin{align}
    \left|\frac{d}{d\tau}\arcsin\left(2\beta(\tau)-1\right)\right|\leq\frac{\sqrt{\mathcal{A}^{\oplus}(\tau)}}{\tau}.
    \label{eq:zetatau_ode}
\end{align}
Using the properties of $N_\circ(\Gamma)$,
$\beta(0)=1$ holds.
Therefore, we obtain
\begin{align}
    \beta(\tau)\geq\cos\left(\frac{1}{2}\int^{\tau}_{0}\frac{\sqrt{\mathcal{A}^{\oplus}(t)}}{t}dt\right)^{2}.
    \label{eq:zetatau_bound}
\end{align}
Similarly to Eq.~\eqref{eq:Bayes_error_alphatau}, the Bayes error for trajectory classification is given by
\begin{align}
    P^{\mathrm{Bayes}}_{\mathrm{err}}\geq\frac{1}{2}-\frac{1}{2}\sqrt{1-4\pi_{0}\pi_{1}\beta(\tau)}.
    \label{eq:Perr_min_zetatau}
\end{align}
Substituting Eq.~\eqref{eq:zetatau_bound} into Eq.~\eqref{eq:Perr_min_zetatau} yields Eq.~\resultII{} for the trajectory classification scenario.

\section{Derivation of an alternative lower bound on the Bayes error\label{sec:error_alt_lowerbound_derivation}}

We derive Eq.~\resultUdaI{} in the main text using the uncertainty relation from Ref.~\cite{Hasegawa:2024:ConcentrationIneqPRL} rather than the Cram\'er-Rao inequality.
First, we consider a Markov process expressed by Eq.~\masterUequation{}.
Let us consider the observable $N_{\varnothing}$, which vanishes for $\Gamma_\varnothing$, i.e., $N_\varnothing(\Gamma_\varnothing)=0$. 
Then, the following relation holds \cite{Hasegawa:2024:ConcentrationIneqPRL}:
\begin{align}
    \frac{\mathrm{Var}[N_{\varnothing}]}{\mathbb{E}[N_{\varnothing}]^{2}}\ge\frac{1}{e^{\mathfrak{a}(0)\tau}-1},
    \label{eq:concentration_transient_KUR}
\end{align}
where $\mathfrak{a}(0)$ is the dynamical activity at time $t=0$.
Consider Eq.~\eqref{eq:concentration_transient_KUR} for the combined Markov process.
Then, we have
\begin{align}
    \frac{\mathrm{Var}[N^{\oplus}_{\varnothing}]}{\mathbb{E}[N^{\oplus}_{\varnothing}]^{2}}&\ge\frac{1}{e^{\left(\mathfrak{a}^{(0)}(0)+\mathfrak{a}^{(1)}(0)\right)\tau}-1}\nonumber\\&=\frac{1}{e^{\mathfrak{a}^{\oplus}(0)\tau}-1}.
    \label{eq:concentration_transient_KUR_combined}
\end{align}
Here, $N_\varnothing^\oplus$ is the observable in the combined Markov process that vanishes for no-jump trajectories, i.e., $N^{\oplus}_{\varnothing}(\Gamma^{\oplus}_{\varnothing})=0$, where $\Gamma^\oplus_\varnothing$ is a no-jump trajectory in the combined process.

Similar to the observable considered in Eq.~\eqref{eq:Oplus_def}, we employ the following observable for $N_\varnothing^\oplus$:
\begin{align}
    \Phi^{\oplus}_{\varnothing}(\Gamma^{\oplus})\equiv\begin{cases}
0 & z^{(0)}(\tau)=z^{(1)}(\tau)\\
1 & \mathrm{otherwise}.
\end{cases}
\label{eq:Phi_oplus_def}
\end{align}
In the main text, we have assumed that the initial state is identical for $Y=0$ and $Y=1$, i.e., $z^{(0)}(0)=z^{(1)}(0)$. Consequently, if no jump occurs in $\Gamma^\oplus$, then $z^{(0)}(\tau)=z^{(1)}(\tau)$, and $\Phi^{\oplus}_{\varnothing}(\Gamma^{\oplus}) = 0$ for $\Gamma_\varnothing^\oplus$. Equation~\eqref{eq:concentration_transient_KUR_combined} therefore holds for $\Phi^{\oplus}_{\varnothing}$, and thus $\Phi_\varnothing^\oplus$ can be used in Eq.~\eqref{eq:concentration_transient_KUR_combined} in place of $N_\varnothing^\oplus$.
The expectation and variance for Eq.~\eqref{eq:Phi_oplus_def} are given by
\begin{align}
    \mathbb{E}\left[\Phi^{\oplus}_{\varnothing}(\Gamma^{\oplus})\right]&=1-\alpha(\tau),\label{eq:Phi_expectation}\\\mathrm{Var}[\Phi^{\oplus}_{\varnothing}(\Gamma^{\oplus})]&=1-\alpha(\tau)-(1-\alpha(\tau))^{2},\label{eq:Phi_variance}
\end{align}
where $\alpha(\tau)$ is defined in Eq.~\eqref{eq:alpha_tau_def}. 
Substituting Eqs.~\eqref{eq:Phi_expectation} and \eqref{eq:Phi_variance} into Eq.~\eqref{eq:concentration_transient_KUR_combined}, we obtain
\begin{align}
    e^{-\mathfrak{a}^{\oplus}(0)\tau}\le\alpha(\tau).
    \label{eq:alphatau_lowerbound}
\end{align}
Substituting Eq.~\eqref{eq:alphatau_lowerbound} into Eq.~\eqref{eq:Bayes_error_alphatau},
we obtain
\begin{align}
    P^{\mathrm{Bayes}}_{\mathrm{err}}&\ge\frac{1}{2}\left(1-\sqrt{1-4\pi_{0}\pi_{1}e^{-\mathfrak{a}^{\oplus}(0)\tau}}\right),
    \label{eq:Perr_min_a0tau}
\end{align}
which is the lower bound on the Bayes error in terms of the dynamical activity.
When we employ the uniform prior distribution $\pi_0=\pi_1=1/2$ in Eq.~\eqref{eq:Perr_min_a0tau}, we obtain Eq.~\resultUdaI{} in the main text.

Next, we consider the trajectory-based classification scenario.
We consider the following observable:
\begin{align}
    \Psi^{\oplus}_{\varnothing}(\Gamma^{\oplus})\equiv\begin{cases}
0 & N_\circ(\Gamma^{(0)})=N_\circ(\Gamma^{(1)})\\
1 & \mathrm{otherwise}.
\end{cases}
\label{eq:Noplus_inv_def}
\end{align}
Clearly, when there is no jump in $\Gamma^\oplus$, 
there is no jump in the first and second processes as well. 
Therefore, $N_\circ(\Gamma^{(0)})=N_\circ(\Gamma^{(1)})=\mathrm{const}$, and thus Eq.~\eqref{eq:Noplus_inv_def} vanishes for the no-jump trajectory $\Gamma^\oplus$. 
Consequently, $\Psi_\varnothing^\oplus$ can be used in Eq.~\eqref{eq:concentration_transient_KUR_combined} in place of $N_\varnothing^\oplus$.
The expectation and variance for Eq.~\eqref{eq:Noplus_inv_def} are given by
\begin{align}
    \mathbb{E}\left[\Psi^{\oplus}_{\varnothing}(\Gamma^{\oplus})\right]&=1-\beta(\tau),\label{eq:Psi_expectation}\\\mathrm{Var}[\Psi^{\oplus}_{\varnothing}(\Gamma^{\oplus})]&=1-\beta(\tau)-(1-\beta(\tau))^{2},
    \label{eq:Psi_variance}
\end{align}
where $\beta(\tau)$ is defined in Eq.~\eqref{eq:zetatau_def}. 
Substituting Eqs.~\eqref{eq:Psi_expectation} and \eqref{eq:Psi_variance} into Eq.~\eqref{eq:concentration_transient_KUR_combined}, we obtain
\begin{align}
    e^{-\mathfrak{a}^{\oplus}(0)\tau}\le\beta(\tau).
    \label{eq:betatau_lowerbound}
\end{align}
Substituting Eq.~\eqref{eq:betatau_lowerbound} into Eq.~\eqref{eq:Perr_min_zetatau},
we obtain Eq.~\eqref{eq:Perr_min_a0tau} for trajectory-based classification.

\section{Classification results}

To show how the obtained bounds behave in practical scenarios, we perform classification tasks. 
First, consider the case in which classification is based on the state of a Markov process at time $\tau$.
In this setting, Eqs.~\resultI{}, \resultII{}, and \resultUdaI{} hold for the classification error.
Here, we focus on a two-state Markov process $\mathfrak{B}=\{1,2\}$.
We randomly generate two transition-rate matrices, corresponding to the classes $y=0$ and $y=1$, denoted by $W^{(0)}$ and $W^{(1)}$, respectively, and let the system evolve according to either matrix. We then consider the task of predicting whether the underlying class is $y=0$ or $y=1$ based on the final state, that is, whether $z(\tau)=1$ or $z(\tau)=2$.

\begin{figure*}
\centering
\includegraphics[width=0.85\linewidth]{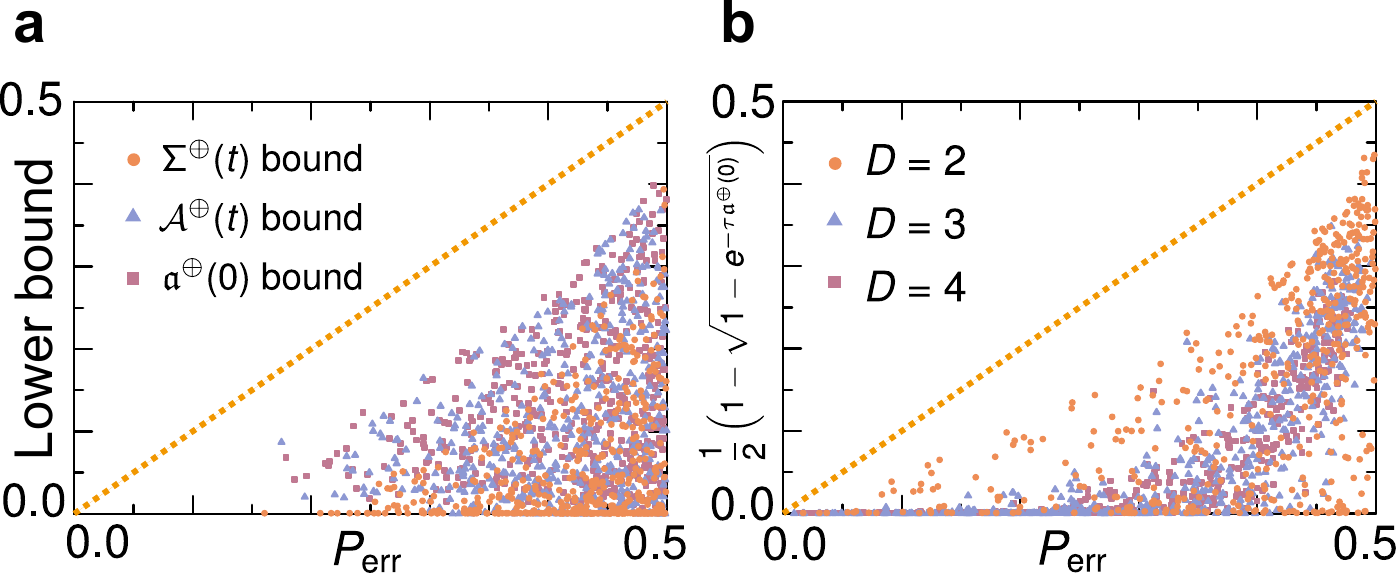}
\caption{
Results of classification. 
(a) State-based classification.
The empirical classification error $P_\mathrm{err}$ is plotted against three lower bounds for random realizations of the Markov process: Eq.~\resultI{} (red circles), Eq.~\resultII{} (blue triangles), and Eq.~\resultUdaI{} (purple squares).
The dashed line indicates the equality case for each bound.
(b) Trajectory-based classification. The classification error $P_\mathrm{err}$ is plotted against the lower bound of Eq.~\resultUdaI{} for Markov processes of three sizes across random realizations: $D=2$ (red circles), $D=3$ (blue triangles), and $D=4$ (purple squares).
The dashed line indicates the equality case of Eq.~\resultUdaI{}.
}
\label{fig:simulation}
\end{figure*}

In Fig.~\ref{fig:simulation}(a), the empirical classification error $P_\mathrm{err}$ and the thermodynamic lower bounds given in Eqs.~\resultI{}, \resultII{}, and \resultUdaI{} are plotted. The red circles represent the lower bound of Eq.~\resultI{} given by the time-integrated entropy production $\Sigma^\oplus(\tau)=\Sigma^{(0)}(\tau)+\Sigma^{(1)}(\tau)$, the blue triangles represent the lower bound of Eq.~\resultII{} given by the time-integrated dynamical activity $\mathcal{A}^\oplus(\tau)=\mathcal{A}^{(0)}(\tau)+\mathcal{A}^{(1)}(\tau)$, and the purple squares represent the lower bound of Eq.~\resultUdaI{} given by the dynamical activity at time $t=0$, $\mathfrak{a}^\oplus(0) = \mathfrak{a}^{(0)}(0)+\mathfrak{a}^{(1)}(0)$.
The dotted lines in Fig.~\ref{fig:simulation}(a) indicate the conditions under which Eqs.~\resultI{}, \resultII{}, and \resultUdaI{} hold with equality.
As can be seen from Fig.~\ref{fig:simulation}(a), all points lie below the dotted lines, showing that all the bounds hold.
Comparing the entropy-production bound [Eq.~\resultI{}] with the other two bounds based on dynamical activity [Eqs.~\resultII{} and \resultUdaI{}], the entropy-production bound is looser than the other two.
Although both Eq.~\resultII{} and Eq.~\resultUdaI{} are bounds based on dynamical activity, it can be seen that there is not much difference between them in terms of tightness.

Next, we consider classification based on trajectory information. 
This scenario is widely observed in classification by biochemical receptors. 
We consider a $D$-state Markov process, and $T(\cdot)$ comprises a logistic classifier. 
We randomly generate the transition-rate matrices corresponding to the classes $y=0$ and $y=1$. 
We generate random training data from the Markov processes for $y=0$ and $y=1$, and we train the logistic regression using the training data.
We consider the steady-state distribution as the initial condition.
Under steady-state conditions, it can be shown that Eq.~\resultUdaI{} is tighter than Eq.~\resultII{} (see the Methods section).
We then calculate the error $P_\mathrm{err}$ for randomly generated test data.
For details on the implementation, see Section~\ref{sec:details_trajectory_logistic}. 

In Fig.~\ref{fig:simulation}(b), the classification error $P_\mathrm{err}$ is plotted against the lower bound given in Eq.~\resultUdaI{}. 
We consider Markov process sizes of $D=2$, $3$, and $4$, which are shown by red circles, blue triangles, and purple rectangles, respectively. 
The dotted lines indicate the conditions under which each bound holds with equality.
As shown in Fig.~\ref{fig:simulation}(b), all points lie below the dotted lines, confirming that all bounds are satisfied.
Comparing different model sizes, it can be seen that smaller models are closer to the dotted line. This indicates that the bound is tighter for smaller model sizes.

\subsection{Details of trajectory-based classification based on logistic regression\label{sec:details_trajectory_logistic}}

We provide details of the trajectory-based classification. 
As shown by Eq.~\trajectoryUsingleUdef{}, each trajectory contains a different number of jumps. Therefore, it is not straightforward to use trajectory information. 
Here, we extract the number of jumps between each state from a trajectory $\Gamma$ and use this information to construct $T(x)=T(\Gamma)$ to obtain the binary prediction. 
Specifically, we use $[N_{\nu\mu}]_{\mu,\nu\,(\mu\ne\nu)}$,
where $N_{\nu\mu}$ is the number of jumps from $\mu$ to $\nu$ during the interval $[0,\tau]$. 
For example, for $D=3$, the feature $[N_{\nu\mu}]_{\mu,\nu\;(\mu\ne \nu)}$ is
\begin{align}
    [N_{\nu\mu}]_{\mu,\nu\;(\mu\ne \nu)}=[N_{21},N_{31},N_{12},N_{32},N_{13},N_{23}].
    \label{eq:X_example}
\end{align}
We use this information to construct $T(\Gamma)$. Specifically, we use logistic regression for $T(\Gamma)$. 
Let us consider the sigmoid function:
\begin{align}
    \mathrm{Sig}([N_{\nu\mu}])=\frac{1}{1+e^{-\theta_{0}-\sum_{\nu,\mu(\nu\ne\mu)}\theta_{\nu\mu}N_{\nu\mu}}}.
    \label{eq:sigmoid_def}
\end{align}
Here, we have abbreviated $[N_{\nu\mu}] \equiv [N_{\nu\mu}]_{\mu,\nu\,(\mu\ne\nu)}$
and 
$\bm{\theta}\equiv[\theta_{0},\theta_{12},\theta_{13}\cdots]$ denotes the set of parameters to be optimized, where $\theta_0$ serves as a bias term.
After optimizing $\bm{\theta}$, the prediction of the label is made by 
\begin{align}
    T(x)=T(\Gamma)=\begin{cases}
1 & \mathrm{Sig}([N_{\nu\mu}])\ge\frac{1}{2}\\
0 & \mathrm{Sig}([N_{\nu\mu}])<\frac{1}{2}
\end{cases}.
\label{eq:logistic_Tx}
\end{align}
In the main text, we have mentioned that Eq.~\resultUdaI{} holds for trajectory-based classification via the observable $N_\circ(\Gamma)$.
$N_\circ(\Gamma)$ is an observable that is constant for no-jump trajectories, $N_\circ(\Gamma_\varnothing)=\mathrm{const}$, where $\Gamma_\varnothing$ is a trajectory with no jumps.
We show that the prediction based on Eq.~\eqref{eq:logistic_Tx} can be represented as classification based on $N_\circ(\Gamma)$.
Let us consider the following observable:
\begin{align}
    N_{\sharp}(\Gamma)=N_{\sharp}([N_{\nu\mu}])=\sum_{\nu,\mu(\nu\ne\mu)}\theta_{\nu\mu}N_{\nu\mu}.
    \label{eq:N_sharp_theta}
\end{align}
By using Eq.~\eqref{eq:N_sharp_theta}, Eq.~\eqref{eq:sigmoid_def} can be expressed as
\begin{align}
    \mathrm{Sig}([N_{\nu\mu}])=\frac{1}{1+e^{-\theta_{0}-N_{\sharp}([N_{\nu\mu}])}}.
    \label{eq:logistic_Nsharp}
\end{align}
Since $N_\sharp$ is a counting observable that vanishes for the no-jump trajectory, it is easy to see that the prediction of Eq.~\eqref{eq:logistic_Tx} is based on an observable that is constant for no-jump trajectories. 
More generally, if a classifier uses the features given in Eq.~\eqref{eq:X_example}, it automatically satisfies the condition required for Eq.~\resultUdaI{}.

Next, we show the implementation details of logistic regression. 
We used logistic regression in scikit-learn \cite{Pedregosa:2011:scikit-learn}. 
To train the logistic regression model, we randomly generate $\mathfrak{N}_{\mathrm{train}}$ samples for the classes $y=0$ and $y=1$. We begin by randomly generating transition-rate matrices $W^{(0)}$ and $W^{(1)}$, then simulate the corresponding Markov processes to produce $\mathfrak{N}_{\mathrm{train}}/2$ training samples for each class.
After training, we calculate the classification error $P_\mathrm{err}$ for $\mathfrak{N}_{\mathrm{pred}}$ randomly generated samples. 
In this study, we employ 
$\mathfrak{N}_{\mathrm{train}}=2000$ and $\mathfrak{N}_{\mathrm{pred}}=2000$.

\section{Comparison of dynamical activity-based bounds\label{sec:bound_comparison}}

In Eqs.~\resultI{} and \resultUdaI{}, we present two bounds involving the dynamical activity. 
In general, it is unclear which bound is tighter. 
Under steady-state conditions, however, Eq.~\resultUdaI{} is tighter, as shown below.
The calculation below is based on Ref.~\cite{Hasegawa:2024:ConcentrationIneqPRL}. 
The inequality to be proved is
\begin{align}
\cos \left[\frac{1}{2} \int_0^\tau \frac{\sqrt{\mathcal{A}(t)}}{t} d t\right]^2 \leq e^{-\mathfrak{a}(0) \tau}.
\label{eq:cos2_expa0tau}
\end{align}
In the steady state, the dynamical activity is given by $\mathcal{A}(\tau)=\mathfrak{a}_{\mathrm{ss}} \tau$, where $\mathfrak{a}_\mathrm{ss}$ is the dynamical activity under steady-state conditions. 
This reduces Eq.~\eqref{eq:cos2_expa0tau} to
\begin{align}
\cos (a) \leq \exp \left[-\frac{a^2}{2}\right],
\label{eq:cosa_expa2}
\end{align}
where $0 \leq a \leq \pi / 2$. Let us define
\begin{align}
R(a) \equiv \ln \cos (a)+\frac{a^2}{2}.
\label{eq:Ra_def}
\end{align}
Since its derivative, $R^{\prime}(a)=a-\tan (a) \leq 0$, for $0 \leq a \leq \pi / 2$, Eq.~\eqref{eq:cos2_expa0tau} holds under steady-state conditions.

\section{Relation to speed limit\label{sec:relation_to_SL}}

Here, we show that weaker bounds in Eqs.~\resultI{} and \resultII{} can be derived via the speed limit. 
We use the speed limit in Refs.~\cite{Hasegawa:2023:BulkBoundaryBoundNC,Nishiyama:2026:TemporalFisherPRE} for consistency with the results.
Let $\mathfrak{p}(x)$ and $\mathfrak{q}(x)$ be two probability distributions, and let $D_B(\mathfrak{p},\mathfrak{q})$ be the Bhattacharyya angle between the distributions:
\begin{align}
    D_B(\mathfrak{p},\mathfrak{q})=\arccos\left(\sum_{x}\sqrt{\mathfrak{p}(x)\mathfrak{q}(x)}\right).
    \label{eq:Bhat_angle_def}
\end{align}
From Ref.~\cite{Hasegawa:2023:BulkBoundaryBoundNC}, the following relation holds:
\begin{align}
    D_{B}(P(\nu;0),P(\nu;\tau))\le\frac{1}{2}\int^{\tau}_{0}\frac{\sqrt{\mathcal{A}(t)}}{t}dt.
    \label{eq:Bhat_angle_SL}
\end{align}
Since $D_B(\cdot,\cdot)$ satisfies the metric axioms, it satisfies the triangle inequality:
\begin{align}
    D_{B}(P^{(0)}(\nu;\tau),P^{(1)}(\nu;\tau))&\le D_{B}(P^{(0)}(\nu;0),P^{(0)}(\nu;\tau))+D_{B}(P^{(1)}(\nu;0),P^{(1)}(\nu;\tau)),
    \label{eq:DB_triangle_inequality}
\end{align}
where we used the relation $P^{(0)}(\nu;0) = P^{(1)}(\nu;0)$. 
Substituting Eq.~\eqref{eq:Bhat_angle_SL} into Eq.~\eqref{eq:DB_triangle_inequality}, we obtain
\begin{align}
    D_{B}(P^{(0)}(\nu;\tau),P^{(1)}(\nu;\tau))&\le\frac{1}{2}\int^{\tau}_{0}\frac{\sqrt{\mathcal{A}^{(0)}(t)}+\sqrt{\mathcal{A}^{(1)}(t)}}{t}dt\nonumber\\&\le\frac{\sqrt{2}}{2}\int^{\tau}_{0}\frac{\sqrt{\mathcal{A}^{(0)}(t)+\mathcal{A}^{(1)}(t)}}{t}dt,
    \label{eq:DB_upperbound_SL}
\end{align}
which yields
\begin{align}
    \sum_{\nu}\sqrt{P^{(0)}(\nu;\tau)P^{(1)}(\nu;\tau)}\ge\cos\left(\frac{\sqrt{2}}{2}\int^{\tau}_{0}\frac{\sqrt{\mathcal{A}^{(0)}(t)+\mathcal{A}^{(1)}(t)}}{t}dt\right).
    \label{eq:final_ineq_SL}
\end{align}
Therefore, the following relation holds:
\begin{align}
    \mathrm{Bhat}\left(P^{(0)}(\nu;\tau),P^{(1)}(\nu;\tau)\right)^2\ge\cos\left(\frac{\sqrt{2}}{2}\int^{\tau}_{0}\frac{\sqrt{\mathcal{A}^{(0)}(t)+\mathcal{A}^{(1)}(t)}}{t}dt\right)^{2}.
    \label{eq:alphatau_upperbound_SL}
\end{align}
In Section~\ref{sec:error_lowerbound_derivation}, the lower bound was 
\begin{align}
    \mathrm{Bhat}\left(P^{(0)}(\nu;\tau),P^{(1)}(\nu;\tau)\right)^2\ge\cos\left(\frac{1}{2}\int^{\tau}_{0}\frac{\sqrt{\mathcal{A}^{(0)}(t)+\mathcal{A}^{(1)}(t)}}{t}dt\right)^{2},
    \label{eq:alphatau_upperbound_It}
\end{align}
which follows directly from Eq.~\eqref{eq:alphatau_bound}.
Since $(\cos x)^2$ is decreasing, Eq.~\eqref{eq:alphatau_upperbound_SL} is weaker than Eq.~\eqref{eq:alphatau_upperbound_It}, indicating that the resulting relation from Eq.~\eqref{eq:alphatau_upperbound_SL} is weaker. 
Therefore, although a similar bound can be derived via the speed limit, it is weaker than the bound derived in Section~\ref{sec:error_lowerbound_derivation}. 
Moreover, the bound in Section~\ref{sec:error_alt_lowerbound_derivation}, which is usually tighter than that in Section~\ref{sec:error_lowerbound_derivation}, cannot be derived via the speed limit.


\begin{table*}
\centering
\begin{tabular}{cccc}
\toprule
Bound & Observable & Thermodynamic quantity & Initial condition\\
\midrule
${\displaystyle  P^{\mathrm{ave}}_{\mathrm{err}}\geq\frac{1}{2}\left[1-\sin\left(\frac{1}{2\sqrt{2}}\int^{\tau}_{0}\frac{\sqrt{\Sigma^{(0)}(t)+\Sigma^{(1)}(t)}}{t}dt\right)\right]}$ & $z(\tau)$ & $\Sigma^{(0)}(t) + \Sigma^{(1)}(t)$ & $z^{(0)}(0)=z^{(1)}(0)$\\
\addlinespace
${\displaystyle P^{\mathrm{ave}}_{\mathrm{err}}\ge\frac{1}{2}\left[1-\sin\left(\frac{1}{2}\int^{\tau}_{0}\frac{\sqrt{\mathcal{A}^{(0)}(t)+\mathcal{A}^{(1)}(t)}}{t}dt\right)\right]}$ & $z(\tau)$ & $\mathcal{A}^{(0)}(t)+\mathcal{A}^{(1)}(t)$ & $z^{(0)}(0)=z^{(1)}(0)$\\
\addlinespace
${\displaystyle P^{\mathrm{ave}}_{\mathrm{err}}\ge\frac{1}{2}\left[1-\sin\left(\frac{1}{2}\int^{\tau}_{0}\frac{\sqrt{\mathcal{A}^{(0)}(t)+\mathcal{A}^{(1)}(t)}}{t}dt\right)\right]}$ & $N_{\circ}(\Gamma)$ & $\mathcal{A}^{(0)}(t)+\mathcal{A}^{(1)}(t)$ & Any\\
\addlinespace
${\displaystyle P^{\mathrm{ave}}_{\mathrm{err}}\ge\frac{1}{2}\left(1-\sqrt{1-e^{-\tau\left[\mathfrak{a}^{(0)}(0)+\mathfrak{a}^{(1)}(0)\right]}}\right)}$ & $z(\tau)$ & $\mathfrak{a}^{(0)}(0)+\mathfrak{a}^{(1)}(0)$ & $z^{(0)}(0)=z^{(1)}(0)$\\
\addlinespace
${\displaystyle P^{\mathrm{ave}}_{\mathrm{err}}\ge\frac{1}{2}\left(1-\sqrt{1-e^{-\tau\left[\mathfrak{a}^{(0)}(0)+\mathfrak{a}^{(1)}(0)\right]}}\right)}$ & $N_{\circ}(\Gamma)$ & $\mathfrak{a}^{(0)}(0)+\mathfrak{a}^{(1)}(0)$ & Any\\
\bottomrule
\end{tabular}

\caption{
Summary of results for the classification of $y=0$ and $1$ (Fig.~\FIGclassificationUscenario{}(a) in the main text).
Lower bounds on the average error $P_\mathrm{err}^{\mathrm{ave}}$ are summarized in terms of the dynamics, observables, thermodynamic quantities, and initial conditions. 
$z(\tau)$ is the state of the Markov process at the final time $t=\tau$.
$\Gamma$ is a trajectory over the interval $[0,\tau]$, and 
$N_\circ(\Gamma)$ is an observable that remains constant along no-jump trajectories.
\label{tab:summary_results}
}
\end{table*}

\begin{table*}
\centering
\begin{tabular}{cccc}
\toprule
Bound & Observable & Thermodynamic quantity & Initial condition\\
\midrule
${\displaystyle  P^{\mathrm{wst}}_{\mathrm{err}}\geq\frac{1}{2}\left[1-\sin\left(\frac{1}{2\sqrt{2}}\int^{\tau}_{0}\frac{\sqrt{\Sigma^{[v]}(t)+\Sigma^{[v^\prime]}(t)}}{t}dt\right)\right]}$ & $z(\tau)$ & $\Sigma^{[v]}(t) + \Sigma^{[v^\prime]}(t)$ & $z^{[v]}(0)=z^{[v^\prime]}(0)$\\
\addlinespace
${\displaystyle P^{\mathrm{wst}}_{\mathrm{err}}\ge\frac{1}{2}\left[1-\sin\left(\frac{1}{2}\int^{\tau}_{0}\frac{\sqrt{\mathcal{A}^{[v]}(t)+\mathcal{A}^{[v^\prime]}(t)}}{t}dt\right)\right]}$ & $z(\tau)$ & $\mathcal{A}^{[v]}(t)+\mathcal{A}^{[v^\prime]}(t)$ & $z^{[v]}(0)=z^{[v^\prime]}(0)$\\
\addlinespace
${\displaystyle P^{\mathrm{wst}}_{\mathrm{err}}\ge\frac{1}{2}\left[1-\sin\left(\frac{1}{2}\int^{\tau}_{0}\frac{\sqrt{\mathcal{A}^{[v]}(t)+\mathcal{A}^{[v^\prime]}(t)}}{t}dt\right)\right]}$ & $N_{\circ}(\Gamma)$ & $\mathcal{A}^{[v]}(t)+\mathcal{A}^{[v^\prime]}(t)$ & Any\\
\addlinespace
${\displaystyle P^{\mathrm{wst}}_{\mathrm{err}}\ge\frac{1}{2}\left(1-\sqrt{1-e^{-\tau\left[\mathfrak{a}^{[v]}(0)+\mathfrak{a}^{[v^\prime]}(0)\right]}}\right)}$ & $z(\tau)$ & $\mathfrak{a}^{[v]}(0)+\mathfrak{a}^{[v^\prime]}(0)$ & $z^{[v]}(0)=z^{[v^\prime]}(0)$\\
\addlinespace
${\displaystyle P^{\mathrm{wst}}_{\mathrm{err}}\ge\frac{1}{2}\left(1-\sqrt{1-e^{-\tau\left[\mathfrak{a}^{[v]}(0)+\mathfrak{a}^{[v^\prime]}(0)\right]}}\right)}$ & $N_{\circ}(\Gamma)$ & $\mathfrak{a}^{[v]}(0)+\mathfrak{a}^{[v^\prime]}(0)$ & Any\\
\bottomrule
\end{tabular}

\caption{
Summary of results for the classification of data points $\mathcal{V}$ (Fig.~\FIGclassificationUscenario{}(b) in the main text). 
Lower bounds on the worst-case error $P_\mathrm{err}^{\mathrm{wst}}$ are summarized in terms of the dynamics, observables, thermodynamic quantities, and initial conditions. 
$v$ and $v^\prime$ in the superscripts should be chosen from $v\in\mathcal{V}_0$ and $v^\prime \in \mathcal{V}_1$. 
$z(\tau)$ is the state of the Markov process at the final time $t=\tau$.
$\Gamma$ is a trajectory over the interval $[0,\tau]$, and 
$N_\circ(\Gamma)$ is an observable that remains constant along no-jump trajectories.
\label{tab:summary_results2}
}
\end{table*}